\documentclass{article}
\usepackage{float}
\usepackage{makeidx}
\usepackage{amssymb}
\usepackage{amsfonts}
\usepackage{amsmath}
\usepackage{graphicx}

\begin{document}

\title{Superconducting Qubits: A Short Review}
\author{M. H. Devoret$^{\dag }$, A. Wallraff$^{\dag }$, and J. M. Martinis$^{\ast }$ \\
{\small $^{\dag}$Department of Applied Physics, Yale University, New Haven, CT 06520}\\
{\small$^{\ast }$Department of Physics, University of California,
Santa Barbara, CA 93106}}\maketitle

\begin{abstract}
Superconducting qubits are solid state electrical circuits
fabricated using techniques borrowed from conventional integrated
circuits. They are based on the Josephson tunnel junction, the
only non-dissipative, strongly non-linear circuit element
available at low temperature. In contrast to microscopic entities
such as spins or atoms, they tend to be well coupled to other
circuits, which make them appealling from the point of view of
readout and gate implementation. Very recently, new designs of
superconducting qubits based on multi-junction circuits have
solved the problem of isolation from unwanted extrinsic
electromagnetic perturbations. We discuss in this review how qubit
decoherence is affected by the intrinsic noise of the junction and
what can be done to improve it.
\end{abstract}

\bigskip


\tableofcontents

\newpage

\section{Introduction}

\subsection{The problem of implementing a quantum computer}

The theory of information has been revolutionized by the discovery
that quantum algorithms can run exponentially faster than their
classical counterparts, and by the invention of quantum
error-correction protocols \cite{Nielsen_Chuang}. \ These
fundamental breakthroughs have lead scientists and engineers to
imagine building entirely novel types of information processors.
However, the construction of a computer exploiting quantum --
rather than classical -- principles represents a formidable
scientific and technological challenge. While quantum bits must be
strongly inter-coupled by gates to perform quantum computation,
they must at the same time be completely decoupled from external
influences, except during the write, control and readout phases
when information must flow freely in and out of the machine. This
difficulty does not exist for the classical bits of an ordinary
computer, which each follow strongly irreversible dynamics that
damp the noise of the environment.

Most proposals for implementing a quantum computer have been based
on qubits constructed from microscopic degrees of freedom: spin of
either electrons or nuclei, transition dipoles of either atoms or
ions in vacuum. These degrees of freedom are naturally very well
isolated from their environment, and hence decohere very slowly.
The main challenge of these implementations is enhancing the
inter-qubit coupling to the level required for fast gate
operations without introducing decoherence from parasitic
environmental modes and noise. \

In this review, we will discuss a radically different experimental
approach based on \textquotedblleft quantum integrated
circuits.\textquotedblright\ \ Here, qubits are constructed from
\textit{collective} electrodynamic modes of macroscopic electrical
elements, rather than microscopic degrees of freedom. \ An
advantage of this approach is that these qubits have intrinsically
large electromagnetic cross-sections, which implies they may be
easily coupled together in complex topologies via simple linear
electrical elements like capacitors, inductors, and transmission
lines. \ However, strong coupling also presents a related
challenge: is it possible to isolate these electrodynamic qubits
from ambient parasitic noise while retaining efficient
communication channels for the write, control, and read
operations? The main purpose of this article is to review the
considerable progress that has been made in the past few years
towards this goal, and to explain how new ideas about methodology
and materials are likely to improve coherence to the threshold
needed for quantum error correction.

\subsection{Caveats}

Before starting our discussion, we must warn the reader that this
review is atypical in that it is neither historical nor
exhaustive. Some important works have not been included or are
only partially covered. We amply cite work by our own, at the risk
of irritating the reader, but we wanted to base our speculations
on experiments whose details we fully understand.  We have on
purpose narrowed our focus: we adopt the point of view of an
engineer trying to determine the best strategy for building a
reliable machine given certain design criteria. This approach
obviously runs the risk of presenting a biased and even incorrect
account of recent scientific results, since the optimization of a
complex system is always an intricate process with both hidden
passageways and dead-ends. We hope nevertheless that the following
sections will at least stimulate discussions on how to harness the
physics of quantum integrated circuits into a mature quantum
information processing technology.

\section{Basic features of quantum integrated circuits}

\subsection{Ultra-low dissipation: superconductivity}

For an integrated circuit to behave quantum mechanically, the
first requirement is the absence of dissipation. More
specifically, all metallic parts need to be made out of a material
that has zero resistance at the qubit operating temperature and at
the qubit transition frequency. This is essential in order for
electronic signals to be carried from one part of the chip to
another without energy loss -- a necessary (but not sufficient)
condition for the preservation of quantum coherence. Low
temperature superconductors such as aluminium or niobium are ideal
for this task \cite{Tinkham}. For this reason, quantum integrated
circuit implementations have been nicknamed \textquotedblleft
superconducting qubits\textquotedblright \footnote{In principle,
other condensed phases of electrons, such as high-Tc
superconductivity or the quantum Hall effect, both integer and
fractional, are possible and would also lead to quantum integrated
circuits of the general type discussed here. We do not pursue this
subject further than this note, however, because dissipation in
these new phases is, by far, not as well understood as in low-Tc
superconductivity.}.

\subsection{Ultra-low noise: low temperature}

The degrees of freedom of the quantum integrated circuit must be
cooled to temperatures where the typical energy $kT$ of thermal
fluctuations is much less that the energy quantum $\hbar \omega
_{01}$ associated with the transition between the states
$\left|qubit=0\right\rangle$ and $\left|qubit=1\right\rangle$. For
reasons which will become clear in subsequent sections, this
frequency for superconducting qubits is in the 5-20 GHz range and
therefore, the operating temperature temperature $T$ must be
around 20 mK (Recall that 1 K corresponds to about 20 GHz). \
These temperatures may be readily obtained by cooling the chip
with a dilution refrigerator. \ Perhaps more importantly though,
the \textquotedblleft electromagnetic
temperature\textquotedblright\ of the wires of the control and
readout ports connected to the chip must also be cooled to these
low temperatures, which requires careful electromagnetic
filtering. \ Note that electromagnetic damping mechanisms are
usually stronger at low temperatures than those originating from
electron-phonon coupling. The techniques
\cite{Martinis-Devoret-Clarke} and requirements \cite{Nahum} for
ultra-low noise filtering have been known for about 20 years. From
the requirements $kT\ll \hbar \omega _{01}$ and $\hbar \omega
_{01}\ll \Delta $, where $\Delta $ is the energy gap of the
superconducting material, one must use superconducting materials
with a transition temperature greater than about 1K. \

\subsection{Non-linear, non-dissipative elements: tunnel junctions}

Quantum signal processing cannot be performed using only purely
linear components. In quantum circuits, however, the non-linear
elements must obey the  additional requirement of being
non-dissipative. Elements like PIN diodes or CMOS transistors are
thus forbidden, even if they could be operated at ultra-low
temperatures.

\begin{figure}[tbp]
\centering
\includegraphics[width = 0.55\columnwidth]{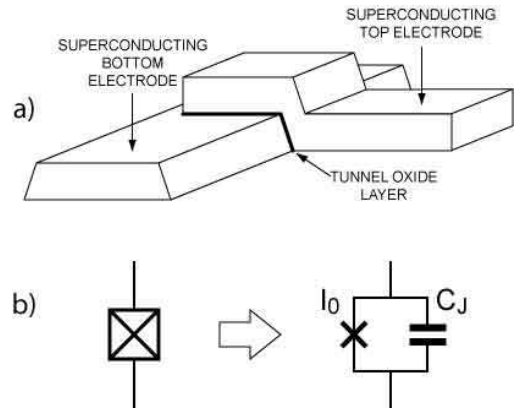}
\caption{a) Josephson tunnel junction made with two
superconducting thin films; b) Schematic representation of a
Josephson tunnel junction. The irreducible Josephson element is
represented by a cross. } \label{fig:fig1-jjab}
\end{figure}

There is only one electronic element that is both non-linear and
non-dissipative at arbitrarily low temperature: the
superconducting tunnel junction\footnote{A very short
superconducting weak link (see for instance Ref.~\cite{Likharev})
is a also a possible candidate, provided the Andreev levels would
be sufficiently separated. Since we have too few experimental
evidence for quantum effects involving this device, we do not
discuss this otherwise important matter further.} (also known as a
Josephson tunnel junction \cite{Josephson-Parks}). As illustrated
in Fig.~\ref{fig:fig1-jjab}, this circuit element consists of a
sandwich of two superconducting thin films separated by an
insulating layer that is thin enough (typically $\sim $1 nm) to
allow tunneling of discrete charges through the barrier. \ In
later sections we will describe how the tunneling of Cooper pairs
creates an inductive path with strong non-linearity, thus creating
energy levels suitable for a qubit. \ The tunnel barrier is
typically fabricated from oxidation of the superconducting metal.
This results in a reliable barrier since the oxidation process is
self-terminating \cite{Giaever}. \ \ The materials properties of
amorphous aluminum oxide, alumina, make it an attractive tunnel
insulating layer. \ In part because of its well-behaved oxide,
aluminum is the material from which good quality  tunnel junctions
are most easily fabricated, and it is often said that aluminium is
to superconducting quantum circuits what silicon is to
conventional MOSFET circuits. Although the Josephson effect is a
subtle physical effect involving a combination of tunneling and
superconductivity, the junction fabrication process is relatively
straightforward.

\subsection{Design and fabrication of quantum integrated circuits}

Superconducting junctions and wires are fabricated using
techniques borrowed from conventional integrated
circuits\footnote{It is worth mentioning that chips with tens of
thousands of junctions have been successfully fabricated for the
voltage standard and for the Josephson signal processors, which
are only exploiting the speed of Josephson elements, not their
quantum properties.}. Quantum circuits are typically made on
silicon wafers using optical or electron-beam lithography and thin
film deposition. They present themselves as a set of micron-size
or sub-micron-size circuit elements (tunnel junctions, capacitors,
and inductors) connected by wires or transmission lines. The size
of the chip and elements are such that, to a large extent, the
electrodynamics of the circuit can be analyzed using simple
transmission line equations or even a lumped element
approximation. Contact to the chip is made by wires bonded to
mm-size metallic pads. \ The circuit can be designed using
conventional layout and classical simulation programs. \

Thus, many of the design concepts and tools of conventional
semiconductor electronics can be directly applied to quantum
circuits. Nevertheless, there are still important differences
between conventional and quantum circuits at the conceptual level.

\subsection{Integrated circuits that obey macroscopic quantum mechanics}

At the conceptual level, conventional and quantum circuits differ
in that, in the former, the collective electronic degrees of
freedom such as currents and voltages are classical variables,
whereas in the latter, these degrees of freedom must be treated by
quantum operators which do not necessarily commute. A more
concrete way of presenting this rather abstract difference is to
say that a typical electrical quantity, such as the charge on the
plates of a capacitor, can be thought of as a simple number in
conventional circuits, whereas in quantum circuits, the charge on
the capacitor must be represented by a wave function giving the
probability amplitude of all charge configurations. For example,
the charge on the capacitor can be in a superposition of states
where the charge is both positive and negative at the same time. \
Similarly the current in a loop might be flowing in two opposite
directions at the same time. These situations have originally been
nicknamed \textquotedblright macroscopic quantum
coherence\textquotedblright effects by Tony Leggett \cite{Leggett}
to emphasize that quantum integrated circuits are displaying
phenomena involving the collective behavior of many particles,
which are in contrast to the usual quantum effects associated with
microscopic particles such as electrons, nuclei or
molecules\footnote{These microscopic effects determine also the
properties of materials, and explain phenomena such as
superconductivity and the Josephson effect itself. Both classical
and quantum circuits share this bottom layer of microscopic
quantum mechanics.}.

\subsection{DiVincenzo criteria}

We conclude this section by briefly mentioning how quantum
integrated circuits satisfy the so-called DiVicenzo criteria for
the implementation of quantum computation \cite{DiVicenzo}. The
non-linearity of tunnel junctions is the key property ensuring
that non-equidistant level subsystems can be implemented
(criterion \# 1: qubit existence). As in many other
implementations, initialization is made possible (criterion \#2:
qubit reset) by the use of low temperature. \ Absence of
dissipation in superconductors is one of the key factors in the
quantum coherence of the system (criterion \# 3: qubit coherence).
Finally, gate operation and readout (criteria \#4 and \#5) are
easily implemented here since electrical signals confined to and
traveling along wires constitute very efficient coupling methods.

\section{The simplest quantum circuit}

\subsection{Quantum LC\ oscillator}

\begin{figure}[tbp]
\centering
\includegraphics[width = 0.20\columnwidth]{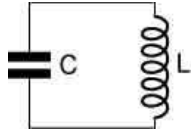}
\caption{Lumped element model for an electromagnetic resonator:
$LC$ oscillator.} \label{fig:fig2-lc}
\end{figure}

We consider first the simplest example of a quantum integrated
circuit, the LC\ oscillator. \ This circuit is shown in
Fig.~\ref{fig:fig2-lc}, and consists of an inductor $L$ connected
to a capacitor $C$, all metallic parts being superconducting. This
simple circuit is the lumped-element version of a superconducting
cavity or a transmission line resonator (for instance, the link
between cavity resonators and $LC$ circuits is elegantly discussed
by Feynman \cite{Feynman}). The equations of motion of the $LC$
circuit are those of an harmonic oscillator. It is convenient to
take the position coordinate as being the flux $\Phi $ in the
inductor, while the role of conjugate momentum is played by the
charge $Q$ on the capacitor playing the role of its conjugate
momentum. The variables $\Phi $ and $Q$ have to be treated as
canonically conjugate quantum operators that obey $\left[ \Phi ,Q
\right] =i\hbar $. The hamiltonian of the circuit is $H=\Phi
^{2}/2L+Q^{2}/2C$, which can be equivalently written as $H=\hbar
\omega _{0}(n+{1}/{2})$ where $n$ is the number operator for
photons in the resonator and $\omega _{0}=1/\sqrt{LC}$ is the
resonance frequency of the oscillator. It is important to note
that the parameters of the circuit hamiltonian are not fundamental
constants of Nature. They are engineered quantities with a large
range of possible values which can be modified easily by changing
the dimensions of elements, a standard lithography operation. It
is in this sense, in our opinion, that the system is unambiguously
\textquotedblleft macroscopic\textquotedblright. The other
important combination of the parameters $L$ and $C$ is the
characteristic impedance $Z=\sqrt{L/C}$ of the circuit. When we
combine this impedance with the residual resistance of the circuit
and/or its radiation losses, both of which we can lump into a
resistance $R$, we obtain the quality factor of the oscillation:
$\mathcal{Q}=R/Z$. The theory of the harmonic oscillator shows
that a quantum superposition of ground state and first excited
state decays on a time scale given by $1/RC$. \ This last equality
illustrates the general link between a classical measure of
dissipation and the upper limit of the quantum coherence time.

\subsection{Practical considerations}

In practice, the circuit shown in Fig.~\ref{fig:fig2-lc} may be
fabricated using planar components with lateral dimensions around
10 $\mu$m, giving values of $L$ and $C$ approximately $0.1$ nH and
$1$ pF, respectively, and yielding $\omega _{0}/2\pi \simeq 16
$GHz and $Z_{0}=10 \, \Omega $. If we use aluminium, a good BCS
superconductor with transition temperature of 1.1 K and a gap
$\Delta /e\simeq$ 200 $\mu V$, dissipation from the breaking of
Cooper pairs will begin at frequencies greater than $2\Delta /h
\simeq 100$GHz. \ The residual resistivity of a BCS superconductor
decreases exponentially with the inverse of temperature and
linearly with frequency, as shown by the Mattis-Bardeen (MB)\
formula $\rho \left( \omega \right) \sim \rho _{0}{\hbar \omega
}/({k_{B}T})\exp \left( -\Delta /k_{B}T\right) $
\cite{Mattis-Bardeen}, where $\rho _{0}$ is the resistivity of the
metal in the normal state (we are treating here the case of the
so-called \textquotedblleft dirty\textquotedblright\
superconductor \cite{DeGennes}, which is well adapted to thin film
systems). According to MB, the intrinsic losses of the
superconductor at the temperature and frequency (typically 20 mK
and 20 GHz) associated with qubit dynamics can be safely
neglected. However, we must warn the reader that the intrisinsic
losses in the superconducting material do not exhaust, by far,
sources of dissipation, even if very high quality factors have
been demonstrated in superconducting cavity experiments
\cite{Haroche-Raimond}.

\subsection{Matching to the vacuum impedance: a useful feature, not a bug}

Although the intrisinsic dissipation of superconducting circuits
can be made very small, losses are in general governed by the
coupling of the circuit with the electromagnetic environment that
is present in the forms of write, control and readout lines. These
lines (which we also refer to as ports) have a characteristic
propagation impedance $Z_{c}\simeq 50\Omega $, which is
constrained to be a fraction of the impedance of the vacuum
$Z_{\mathrm{vac}}=377\Omega $. It is thus easy to see that our
$LC$ circuit, with a characteristic impedance of $Z_{0}=10\Omega
$, tends to be rather well impedance-matched to any pair of leads.
This circumstance occurs very frequently in circuits, and almost
never in microscopic systems such as atoms which interact very
weakly with electromagnetic radiation\footnote{The impedance of an
atom can be crudely seen as being given by the impedance quantum
$R_{K}=h/e^{2}$. We live in a universe where the ratio
$Z_{vac}/2R_{K}$, also known as the fine structure constant
1/137.0, is a small number.}. Matching to $Z_{\mathrm{vac}}$ is a
useful feature because it allows strong coupling for writing,
reading, and logic operations. As we mentioned earlier, the
challenge with quantum circuits is to isolate them from parasitic
degrees of freedom. \textbf{The major task of this review is to
explain how this has been achieved so far and what level of
isolation is attainable.}

\subsection{The consequences of being macroscopic}

While our example shows that quantum circuits can be mass-produced
by standard microfabrication techniques and that their parameters
can be easily engineered to reach some optimal condition, it also
points out evident drawbacks of being \textquotedblleft
macroscopic\textquotedblright\ for qubits.

The engineered quantities $L$ and $C$ can be written as

\begin{eqnarray}
L &=&L^{\rm{stat}}+\Delta L\left( t\right)  \label{fluctuat-LC} \\
C &=&C^{\rm{stat}}+\Delta C\left( t\right)  \notag
\end{eqnarray}

\begin{itemize}
    \item[a)] The first term on the right-handside denotes the static part of
the parameter. It has \textbf{statistical variations}: unlike
atoms whose transition frequencies in isolation are so
reproducible that they are the basis of atomic clocks, circuits
will always be subject to parameter variations from one
fabrication batch to another. Thus prior to any operation using
the circuit, the transition frequencies and coupling strength will
have to be determined by \textquotedblleft
diagnostic\textquotedblright\ sequences and then taken into
account in the algorithms.
    \item[b)] The second term on the right-handside denotes the
time-dependent fluctuations of the parameter. It describes
\textbf{noise }due to residual material defects moving in the
material of the substrate or in the material of the circuit
elements themselves. This noise can affect for instance the
dielectric constant of a capacitor. The low frequency components
of the noise will make the resonance frequency wobble and
contribute to the dephasing of the oscillation. Furthermore, the
frequency component of the noise at the transition frequency of
the resonator will induce transitions between states and will
therefore contribute to the energy relaxation.
\end{itemize}

Let us stress that statistical variations and noise are not
problems affecting superconducting qubit parameters only. For
instance when several atoms or ions are put together in
microcavities for gate operation, patch potential effects will
lead to expressions similar in form to Eq.~(\ref{fluctuat-LC}) for
the parameters of the hamiltonian, even if the isolated single
qubit parameters are fluctuation-free.

\subsection{The need for non-linear elements}

Not all aspects of quantum information processing using quantum
integrated circuits can be discussed within the framework of the
LC circuit, however. It lacks an important ingredient:
non-linearity. In the harmonic oscillator, all transitions between
neighbouring states are degenerate as a result of the parabolic
shape of the potential. In order to have a qubit, the transition
frequency between states $\left|\rm{qubit}=0\right\rangle$ and
$\left|\rm{qubit}=1\right\rangle$ must be sufficiently different
from the transition between higher-lying eigenstates, in
particular 1 and 2. Indeed, the maximum number of 1-qubit
operations that can be performed coherently scales as
$\mathcal{Q}_{01}\left\vert \omega _{01}-\omega _{12}\right\vert
/\omega _{01}$ where $\mathcal{Q}_{01}$ is the quality factor of
the $0\rightarrow 1$ transition. Josephson tunnel junctions are
crucial for quantum circuits since they bring a strongly
non-parabolic inductive potential energy.

\section{The Josephson non-linear inductance}

At low temperatures, and at the low voltages and low frequencies
corresponding to quantum information manipulation, the Josephson
tunnel junction behaves as a pure non-linear inductance (Josephson
element) in parallel with the capacitance corresponding to the
parallel plate capacitor formed by the two overlapping films of
the junction (Fig.~\ref{fig:fig1-jjab}b). This minimal, yet
precise model, allows arbitrary complex quantum circuits to be
analysed by a quantum version of conventional circuit theory. Even
though the tunnel barrier is a layer of order ten atoms thick, the
value of the Josephson non-linear inductance is very robust
against static disorder, just like an ordinary inductance -- such
as the one considered in section 3 -- is very insensitive to the
position of each atom in the wire. We refer to
\cite{Martinis-LesHouches} for a detailed discussion of this
point.

\subsection{Constitutive equation}

Let us recall that a linear inductor, like any electrical element,
can be fully characterized by its constitutive equation.
Introducing a generalization of the ordinary magnetic flux, which
is only defined for a loop, we define the \textbf{branch flux of
an electric element} by $\Phi (t)=\int_{-\infty
}^{t}V(t_{1})dt_{1}$, where $V\left( t\right) $ is the space
integral of the electric field along a current line inside the
element. In this language, the current $I(t)$ flowing through the
inductor is proportional to its branch flux $\Phi (t)$:

\begin{equation}
I\left( t\right) =\frac{1}{L}\Phi (t)  \label{inductor}
\end{equation}

Note that the generalized flux $\Phi (t)$ can be defined for any
electric element with two leads (dipole element), and in
particular for the Josephson junction, even though it does not
resemble a coil. The Josephson element behaves inductively, as its
branch flux-current relationship \cite{Josephson-Parks} is:

\begin{equation}
I\left( t\right) =I_{0}\sin \left[ 2\pi \Phi (t)/\Phi _{0}\right]
\label{nl-inductor}
\end{equation}

\begin{figure}[tbp]
\centering
\includegraphics[width = 0.6\columnwidth]{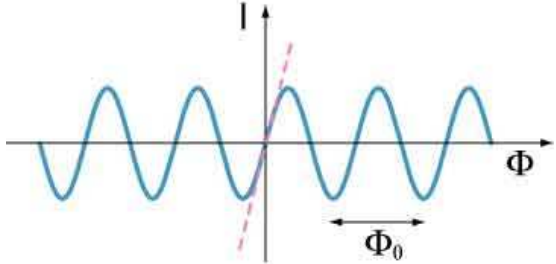}
\caption{Sinusoidal current-flux relationship of a Josephson
tunnel junction, the simplest non-linear, non-dissipative
electrical element (solid line). Dashed line represents
current-flux relationship for a linear inductance equal to the
junction effective inductance. } \label{fig:fig3-iphi}
\end{figure}

This inductive behavior is the manifestation, at the level of
collective electrical variables, of the inertia of Cooper pairs
tunneling across the insulator (kinetic inductance). The
discreteness of Cooper pair tunneling causes the periodic flux
dependence of the current, with a period given by a universal
quantum constant $\Phi _{0}$, the superconducting flux quantum
$h/2e$. The junction parameter $I_{0}$ is called the critical
current of the tunnel element. It scales proportionally to the
area of the tunnel layer and diminishes exponentially with the
tunnel layer thickness. Note that the constitutive relation
Eq.~(\ref{nl-inductor}) expresses in only one equation the two
Josephson relations \cite{Josephson-Parks}. This compact
formulation is made possible by the introduction of the branch
flux (see Fig.~\ref{fig:fig3-iphi}).

The purely sinusoidal form of the constitutive relation
Eq.~(\ref{nl-inductor}) can be traced to the perturbative nature
of Cooper pair tunneling in a tunnel junction. Higher harmonics
can appear if the tunnel layer becomes very thin, though their
presence would not fundamentally change the discussion presented
in this review. The quantity $2\pi \Phi (t)/\Phi _{0}=\delta $ is
called the gauge-invariant phase difference accross the junction
(often abridged into \textquotedblleft phase\textquotedblright ).
It is important to realize that at the level of the constitutive
relation of the Josephson element, this variable is nothing else
than an electromagnetic flux in dimensionless units. In general,
we have

\begin{equation*}
\theta =\delta \, {\rm{mod}}2\pi
\end{equation*}
where $\theta $ is the phase difference between the two
superconducting condensates on both sides of the junction. This
last relation expresses how the superconducting ground state and
electromagnetism are tied together.

\subsection{Other forms of the parameter describing the Josephson non-linear
inductance}

The Josephson element is also often described by two other
parameters, each of which carry exactly the same information as
the critical current. The first one is the Josephson effective
inductance $L_{J0}=\varphi _{0}/I_{0}$, where $\varphi _{0}=\Phi
_{0}/2\pi $ is the reduced flux quantum. The name of this other
form becomes obvious if we expand the sine function in
Eq.~(\ref{nl-inductor}) in powers of $\Phi $ around $\Phi =0$.
Keeping the leading term, we have $I=\Phi /L_{J0}$. Note that the
junction behaves for small signals almost as a point-like kinetic
inductance: a 100nm$\times $100nm area junction will have a
typical inductance of 100nH, whereas the same inductance is only
obtained magnetically with a loop of about 1cm in diameter. More
generally, it is convenient to define the phase-dependent
Josephson inductance

\begin{equation*}
L_{J}\left( \delta \right) =\left( \frac{\partial I}{\partial \Phi
}\right) ^{-1}=\frac{L_{J0}}{\cos \delta }
\end{equation*}

Note that the Josephson inductance not only depends on $\delta $,
it can actually become infinite or negative! Thus, under the
proper conditions, the Josephson element can become a switch and
even an active circuit element, as we will see below.

The other useful parameter is the Josephson energy $E_{J}=\varphi
_{0}I_{0}$. If we compute the energy stored in the junction
$E(t)=\int_{-\infty }^{t}I\left( t_{1}\right) V\left( t_{1}\right)
dt_{1}$, we find $E(t)=-E_{J}\cos \left[ 2\pi \Phi (t)/\Phi
_{0}\right] $. In contrast with the parabolic dependence on flux
of the energy of an inductance, the potential associated with a
Josephson element has the shape of a cosine washboard. The total
height of the corrugation of the washboard is $2E_{J}$.

\subsection{Tuning the Josephson element}

A direct application of the non-linear inductance of the Josephson
element is obtained by splitting a junction and its leads into 2
equal junctions, such that the resulting loop has an inductance
much smaller the Josephson inductance. The two smaller junctions
in parallel then behave as an effective junction \cite{Clarke}
whose Josephson energy varies with $\Phi _{\rm{ext}}$, the
magnetic flux externally imposed through the loop:
\begin{equation}
E_{J}\left( \Phi _{\rm{ext}}\right) =E_{J}\cos \left( \pi \Phi
_{\rm{ext}}/\Phi _{0}\right)   \label{J-flux}
\end{equation}
Here, $E_{J}$ the total Josephson energy of the two junctions. The
Josephson energy can also be modulated by applying a magnetic
field in the plane parallel to the tunnel layer.


\section{The quantum isolated Josephson junction}

\subsection{Form of the hamiltonian}

If we leave the leads of a Josephson junction unconnected, we
obtain the simplest example of a non-linear electrical resonator.
In order to analyze its quantum dynamics, we apply the
prescriptions of quantum circuit theory briefly summarized in
Appendix~\ref{app:quantumcircuittheory}. Choosing a representation
privileging the branch variables of the Josephson element, the
momentum corresponds to the charge $Q=2eN$ having tunneled through
the element and the canonically conjugate position is the flux
$\Phi =\varphi _{0}\theta $ associated with the superconducting
phase difference across the tunnel layer. Here, $N$ and $\theta$
are treated as operators that obey $\left[ \theta ,N\right] =i$.
It is important to note that the operator $N$ has integer
eigenvalues whereas the phase $\theta $ is an operator
corresponding to the position of a point on the unit circle (an
angle modulo $2\pi $).

By eliminating the branch charge of the capacitor, the hamiltonian
reduces to

\begin{equation}
H=E_{CJ}\left( N-Q_{r}/2e\right) ^{2}-E_{J}\cos \theta
\label{single-junction}
\end{equation}
where $E_{CJ}=\frac{(2e)^{2}}{2C_{J}}$ is the Coulomb charging
energy corresponding to one Cooper pair on the junction
capacitance $C_{J}$ and where $Q_{r}$ is the residual offset
charge on the capacitor.

One may wonder how the constant $Q_{r}$ got into the hamiltonian,
since no such term appeared in the corresponding $LC$ circuit in
section 3. The continuous charge $Q_{r}$ is equal to the charge
that pre-existed on the capacitor when it was wired with the
inductor. Such offset charge is not some nit-picking theoretical
construct. Its physical origin is a slight difference in work
function between the two electrodes of the capacitor and/or an
excess of charged impurities in the vicinity of one of the
capacitor plates relative to the other. The value of $Q_{r}$ is in
practice very large compared to the Cooper pair charge $2e$, and
since the hamiltonian~(\ref{single-junction}) is invariant under
the transformation $N\rightarrow N\pm 1$, its value can be
considered completely random.

Such residual offset charge also exists  in the LC circuit.
However, we did not include it in our description of section 3
since a time-independent $Q_{r}$ does not appear in the dynamical
behavior of the circuit: it can be removed from the hamiltonian by
performing a trivial canonical transformation leaving the form of
the hamiltonian unchanged.

It is not possible, however, to iron this constant out of the
junction hamiltonian~(\ref{single-junction}) because the potential
is not quadratic in $\theta$. The parameter $Q_{r}$ plays a role
here similar to the vector potential appearing in the hamiltonian
of an electron in a magnetic field.

\subsection{Fluctuations of the parameters of the hamiltonian}

The hamiltonian~(\ref{single-junction}) thus depends thus on three
parameters which, following our discussion of the $LC$ oscillator,
we write as

\begin{eqnarray}
Q_{r} &=&Q_{r}^{\rm{stat}}+\Delta Q_{r}\left( t\right)
\label{intrinsic-noises}
\\
E_{C} &=&E_{C}^{\rm{stat}}+\Delta E_{C}\left( t\right)   \notag \\
E_{J} &=&E_{J}^{\rm{stat}}+\Delta E_{J}\left( t\right)   \notag
\end{eqnarray}
in order to distinguish the static variation resulting from
fabrication  of the circuit from the time-dependent fluctuations.
While $Q_{r}^{\rm{stat}}$ can be considered fully random (see
above discussion), $E_{C}^{\rm{stat}}$ and $E_{J}^{\rm{stat}}$ can
generally be adjusted by construction to a precision better than
20\%. The relative fluctuations $\Delta Q_{r}\left( t\right) /2e$
and $\Delta E_{J}\left( t\right) /E_{J}$ are found to have a $1/f$
power spectral density with a typical standard deviations at 1 Hz
roughly of order 10$^{-3}$ Hz$^{-1/2}$ and 10$^{-5}$ Hz$^{-1/2}$,
respectively, for a junction with a typical area of 0.01 $\mu
\mathrm{m}^{2}$ \cite{Van-Harlingen}. The noise appears to be
produced by independent two-level fluctuators \cite{Martinis-TLS}.
The relative fluctuations $\Delta E_{C}\left( t\right) /E_{C} $
are much less known, but the behavior of some glassy insulators at
low temperatures might lead us to expect also a $1/f$ power
spectral density, but probably with a weaker intensity than those
of $\Delta E_{J}\left( t\right) /E_{J}$. We refer to the 3 noise
terms in Eq.~(\ref{intrinsic-noises}) as offset charge, dielectric
and critical current noises, respectively.

\section{Why three basic types of Josephson qubits?}

The first-order problem in realizing a Josephson qubit is to
suppress as much as possible the detrimental effect of the
fluctuations of $Q_{r}$, while retaining the non-linearity of the
circuit. There are three main stategies for solving this problem
and they lead to three fundamental basic type of qubits involving
only one Josephson element.

\subsection{The Cooper pair box}

The simplest circuit is called the \textquotedblleft Cooper pair
box\textquotedblright\ and was first described theoretically,
albeit in a slightly different version than presented here, by M.
B\"{u}ttiker \cite{Buttiker}. It was first realized experimentally
by the Saclay group in 1997 \cite{Bouchiat-etal}. Quantum dynamics
in the time domain were first seen by the NEC\ group in 1999
\cite{Nakamura-1stRabi}.

\begin{figure}[!h]
\centering
\includegraphics[width = 0.75\columnwidth]{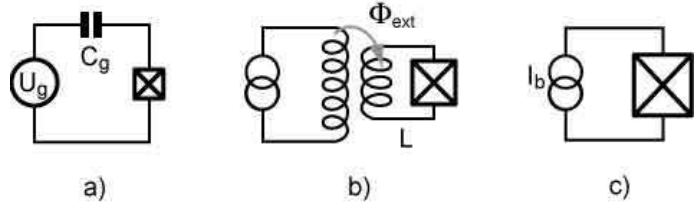}
\caption{The three basic superconducting qubits. a) Cooper pair
box (prototypal charge qubit), b) RF-SQUID (prototypal flux qubit)
and c) current-biased junction (prototypal phase qubit). The
charge qubit and the flux qubit requires small junctions
fabricated with e-beam lithography while the phase qubit can be
fabricated with conventional optical lithography.}
\label{fig:fig4-3qbts}
\end{figure}

In the Cooper pair box, the deviations of the residual offset
charge $Q_{r}$ are compensated  by biasing the Josephson tunnel
junction with a voltage source $U$ in series with a
\textquotedblleft gate\textquotedblright\ capacitor $C_{g}$ (see
Fig.~\ref{fig:fig4-3qbts}a). One can easily show that the
hamiltonian of the Cooper pair box is
\begin{equation}
H=E_{C}\left( N-N_{g}\right) ^{2}-E_{J}\cos \theta \label{ham-CPB}
\end{equation}
Here $E_{C}={\left( 2e\right) ^{2}}/\left({2\left(
C_{J}+C_{g}\right)}\right)$ is the charging energy of the island
of the box and $N_{g}=Q_{r}+C_{g}U/2e$. Note that this hamiltonian
has the same form as hamiltonian~(\ref{single-junction}). Often
$N_{g}$ is simply written as $C_{g}U/2e$ since $U$ at the chip
level will anyway deviate substantially from the generator value
at high-temperature due to stray emf's in the low-temperature
cryogenic wiring.

\begin{figure}[!b]
\centering
\includegraphics[width = 0.6\columnwidth]{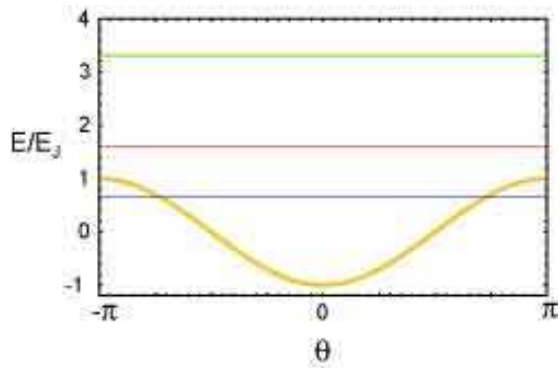}
\caption{Potential landscape for the phase in a Cooper pair box
(thick solid line). The first few levels for $E_{J}/E_{C}=1$ and
$N_{g}=1/2$ are indicated by thin horizontal solid lines.}
\label{fig:fig5-box}
\end{figure}

\begin{figure}[tbp]
\centering
\includegraphics[width = 0.7\columnwidth]{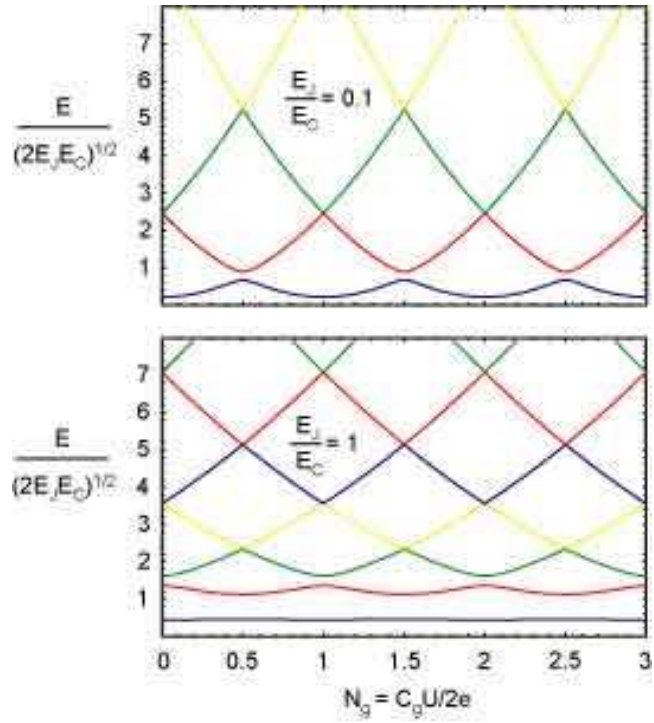}
\caption{Energy levels of the Cooper pair box as a function of
$N_{g}$, for two values of $E_{J}/E_{C}$. As $E_{J}/E_{C}$
increases, the sensitivity of the box to variations of offset
charge diminishes, but so does the non-linearity. However, the
non-linearity is the slowest function of $E_{J}/E_{C}$ and a
compromise advantageous for coherence can be found.}
\label{fig:fig6-2energies-vs-ng}
\end{figure}

In Fig.~\ref{fig:fig5-box} we show the potential in the $\theta $
representation as well as the first few energy levels for
$E_{J}/E_{C}=1$ and $N_{g}=0$. As shown in
Appendix~\ref{app:CPBeigenenergies}, the Cooper pair box
eigenenergies and eigenfunctions can be calculated with special
functions known with arbitrary precision, and in
Fig.~\ref{fig:fig6-2energies-vs-ng} we plot the first few
eigenenergies as a function of $N_{g}$ for $E_{J}/E_{C}=0.1$ and
$E_{J}/E_{C}=1$. Thus, the Cooper box is to quantum circuit
physics what the hydrogen atom is to atomic physics. We can modify
the spectrum with the action of two externally controllable
electrodynamic parameters: $N_{g}$, which is directly proportional
to $U$, and $E_{J}$, which can be varied by applying a field
through the junction or by using a split junction and applying a
flux through the loop, as discussed in section 3. These parameters
bear some resemblance to the Stark and Zeeman fields in atomic
physics. For the box, however much smaller values of the fields
are required to change the spectrum entirely.

We now limit ourselves to the two lowest levels of the box. Near
the degeneracy point $N_{g}=1/2$ where the electrostatic energy of
the of the two charge states $\left\vert N=0\right\rangle $ and
$\left\vert N=1\right\rangle $ are equal, we get the reduced
hamiltonian \cite{Bouchiat-etal, Schoen}

\begin{equation}
H_{\rm{qubit}}=-E_{z}\left( \sigma _{Z}+X_{\rm{control}}\sigma
_{X}\right) \label{box-qubit-ham}
\end{equation}
where, in the limit $E_{J}/E_{C}\ll 1$, $E_{z}={E_{J}}/{2}$and
$X_{\rm{control}}=2{E_{C}}/{E_{J}}\left({1}/{2}-N_{g}\right) $. In
Eq.~(\ref{box-qubit-ham}), $\sigma _{Z}$ and $\sigma _{X}$ refer
to the Pauli spin operators. Note that the $\ X$ direction is
chosen along the charge operator, the variable of the box we can
naturally couple to.

\begin{figure}[!t]
\centering
\includegraphics[width = 0.7\columnwidth]{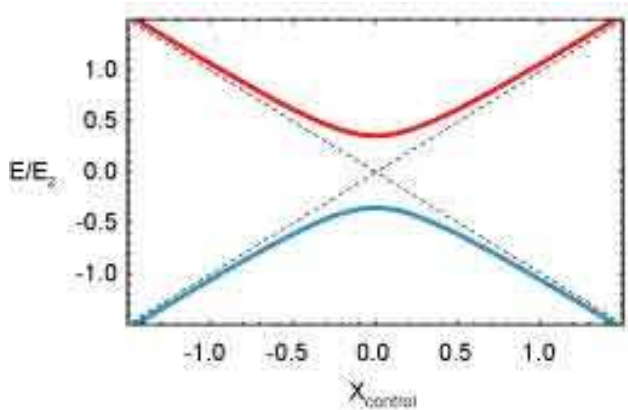}
\caption{Universal level anticrossing found both for the Cooper
pair box and the RF-SQUID at their \textquotedblleft sweet
spot\textquotedblright.} \label{fig:fig7-avoided-crossing-xlabel}
\end{figure}

If we plot the energy of the eigenstates of~(\ref{box-qubit-ham})
as a function of the control parameter $X_{\rm{control}}$, we
obtain the universal level repulsion diagram shown in
Fig.~\ref{fig:fig7-avoided-crossing-xlabel}. Note that the minimum
energy splitting is given by $E_{J}$. Comparing
Eq.~(\ref{box-qubit-ham}) with the spin hamiltonian in NMR, we see
that $E_{J}$ plays the role of the Zeeman field while the
electrostatic energy plays the role of the transverse field.
Indeed we can send on the control port corresponding to $U$
time-varying voltage signals in the form of NMR-type pulses and
prepare arbitrary superpositions of states \cite{Vion-Science}.

The expression~(\ref{box-qubit-ham}) shows that at the
\textquotedblleft sweet spot\textquotedblright\
$X_{\rm{control}}=0$, i.e. the degeneracy point $N_{g}={1}/{2}$,
the qubit transition frequency is to first order insentive to the
offset charge noise $\Delta Q_{r}$. We will discuss in the next
section how an extension of the Cooper pair box circuit can
display quantum coherence properties on long time scales by using
this property.

In general, circuits derived from the Cooper pair box have been
nicknamed \textquotedblleft charge qubits\textquotedblright . One
should not think, however, that in charge qubits, quantum
information is \textit{encoded }with charge. Both the charge $N$
and phase $\theta $ are quantum variables and they are both
uncertain for a generic quantum state. Charge in \textquotedblleft
charge qubits\textquotedblright\ should be understood as refering
to the "controlled variable", i.e. the qubit variable that couples
to the control line we use to write or manipulate quantum
information. In the following, for better comparison between the
three qubits, we will be faithful to the convention used in
Eq.~(\ref{box-qubit-ham}), namely that $\sigma _{X}$ represents
the $\emph{controlled~variable}$.

\subsection{The RF-SQUID}

The second circuit -- the so-called RF-SQUID \cite{Barone-Paterno}
-- can be considered in several ways the dual of the Cooper pair
box (see Fig.~\ref{fig:fig4-3qbts}b). It employs a superconducting
transformer rather than a gate capacitor to adjust the
hamiltonian. The two sides of the junction with capacitance
$C_{J}$ are connected by a superconducting loop with inductance
$L$. An external flux $\Phi _{\rm{ext}}$ is imposed through the
loop by an auxiliary coil. Using the methods of
Appendix~\ref{app:quantumcircuittheory}, we obtain the hamiltonian
\cite{Leggett}

\begin{equation}
H=\frac{q^{2}}{2C_{J}}+\frac{\phi ^{2}}{2L}-E_{J}\cos \left[
\frac{2e}{\hbar }\left( \phi -\Phi _{\rm{ext}}\right) \right]
\label{rfham}
\end{equation}

We are taking here as degrees of freedom the integral $\phi $ of
the voltage across the inductance $L$, i.e. the flux through the
superconducting loop, and its conjugate variable, the charge $q$
on the capacitance $C_{J}$; they obey $\left[ \phi ,q\right]
=i\hbar $. Note that in this representation, the phase $\theta $,
corresponding to the branch flux across the Josephson element, has
been eliminated. Note also that the flux $\phi $, in contrast to
the phase $\theta $, takes its values on a line and not on a
circle. Likewise, its conjugate variable $q$, the charge on the
capacitance, has continuous eigenvalues and not integer ones like
$N$. Note that we now have three adjustable energy scales:\
$E_{J}$, $E_{CJ}={(2e)^{2}}/{2C_{J}}$ and $E_{L}={\Phi
_{0}^{2}}/{2L}$.

\begin{figure}[tbp]
\centering
\includegraphics[width = 0.7\columnwidth]{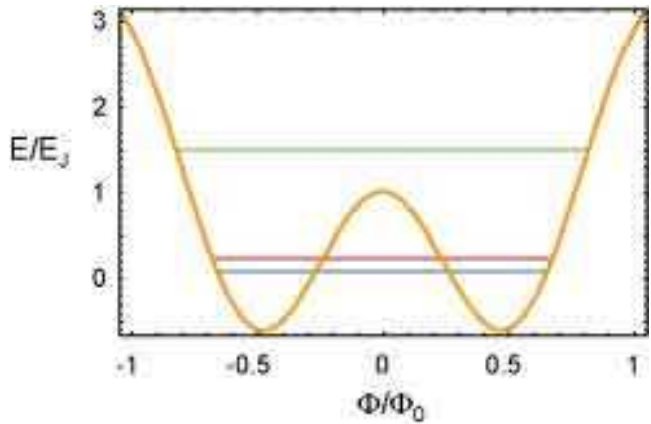}
\caption{Schematic potential energy landcape for the RF-SQUID.}
\label{fig:fig8-rf-squid}
\end{figure}

The potential in the flux representation is schematically shown in
Fig.~\ref{fig:fig8-rf-squid} together with the first few levels,
which have been seen experimentally for the first time by the SUNY
group \cite{Lukens}. Here, no analytical expressions exist for the
eigenvalues and the eigenfunctions of the problem, which has two
aspect ratios: $E_{J}/E_{CJ}$ and $\lambda =L_{J}/L-1$.

Whereas in the Cooper box the potential is cosine-shaped and has
only one well since the variable $\theta $ is $2\pi $-periodic, we
have now in general a parabolic potential with a cosine
corrugation. The idea here for curing the detrimental effect of
the offset charge fluctuations is very different than in the box.
First of all $Q_{r}^{\rm{stat}}$ has been neutralized by shunting
the 2 metallic electrodes of the junction by the superconducting
wire of the loop. Then, the ratio $E_{J}/E_{CJ}$ is chosen to be
much larger than unity. This tends to increase the relative
strength of quantum fluctuations of $q$, making offset charge
fluctuations $\Delta Q_{r}$ small in comparison. The resulting
loss in the non-linearity of the first levels is compensated by
taking $\lambda $ close to zero and by flux-biasing the device at
the half-flux quantum value $\Phi _{\rm{ext}}=\Phi _{0}/2$. Under
these conditions, the potential has two degenerate wells separated
by a shallow barrier with height $E_{B}=E_{J}{3\lambda ^{2}}/{2}$.
This corresponds to the degeneracy value $N_{g}=1/2$ in the Cooper
box, with the inductance energy in place of the capacitance
energy. At $\Phi _{\rm{ext}}=\Phi _{0}/2$, the two lowest energy
levels are then the symmetric and antisymmetric combinations of
the two wavefunctions localized in each well, and the energy
splitting between the two states can be seen as the tunnel
splitting associated with the quantum motion through the potential
barrier between the two wells, bearing close resemblance to the
dynamics of the ammonia molecule. This splitting $E_{S}$ depends
exponentially on the barrier height, which itself depends strongly
on $E_{J}$. We have $E_{S}=\eta \sqrt{ E_{B}E_{CJ}}\exp \left(
-\xi \sqrt{E_{B}/E_{CJ}}\right) $ where the numbers $\eta $ and
$\xi $ have to be determined numerically in most practical cases.
The non-linearity of the first levels results thus from a subtle
cancellation between two inductances: the superconducting loop
inductance $L$ and the junction effective inductance -$L_{J0}$
which is opposed to $L$ near $\Phi _{\rm{ext}}=\Phi _{0}/2$.
However, as we move away from the degeneracy point $\Phi
_{\rm{ext}}=\Phi _{0}/2$, the splitting $2E_{\Phi }$ between the
first two energy levels varies linearly with the applied flux
$E_{\Phi }=\zeta E_{\rm{L}} \left( N_{\Phi }-1/2\right) $. Here
the parameter $N_{\Phi }=\Phi _{\rm{ext}}/\Phi _{0}$, also called
the flux frustration, plays the role of the reduced gate charge
$N_{g}$. The coefficient $\zeta $ has also to be determined
numerically. We are therefore again, in the vicinity of the flux
degeneracy point $\Phi _{\rm{ext}}=\Phi _{0}/2$ and for
$E_{J}/E_{CJ}\gg 1$, in presence of the universal level repulsion
behavior (see Fig.~\ref{fig:fig7-avoided-crossing-xlabel}) and the
qubit hamiltonian is again given by

\begin{equation}
H_{qubit}=-E_{z}\left( \sigma _{Z}+X_{\rm{control}}\sigma
_{X}\right) \label{SQUID-qubit-ham}
\end{equation}
where now $E_{z}=E_{S}/2$ and $X_{\rm{control}}=2{\zeta
E_{\rm{L}}}/{E_{S}}\left( {1}/{2}-N_{\Phi }\right) $. The qubits
derived from this basic circuit \cite{Mooij-Wal, Mooij-Nak} have
been nicknamed \textquotedblleft flux qubits\textquotedblright .
Again, quantum information is not directly represented here by the
flux $\phi $, which is as uncertain for a general qubit state as
the charge $q$ on the capacitor plates of the junction. The flux
$\phi $ is the system variable to which we couple when we write or
control information in the qubit, which is done by sending current
pulses on the primary of the RF-SQUID transformer, thereby
modulating $N_{\Phi }$, which itself determines the strength of
the pseudo-field in the $X$ direction in the
hamiltonian~(\ref{SQUID-qubit-ham}). Note that the parameters
$E_{S}$, $E_{\Phi }$, and $N_{\Phi }$ are all influenced to some
degree by the critical current noise, the dielectric noise and the
charge noise. Another independent noise can also be present, the
noise of the flux in the loop, which is not found in the box and
which will affect only $N_{\Phi }$. Experiments on DC-SQUIDS
\cite{Clarke} have shown that this noise, in adequate conditions,
can be as low as $10^{-8}(h/2e)\mathrm{Hz}^{-1/2}$ at a few kHz.
However, experimental results on flux qubits (see below) seem to
indicate that larger apparent flux fluctuations are present,
either as a result of flux trapping or critical current
fluctuations in junctions implementing inductances.

\subsection{Current-biased junction}

The third basic quantum circuit biases the junction with a fixed
DC-current source (Fig.~\ref{fig:fig7-avoided-crossing-xlabel}c).
Like the flux qubit, this circuit is also insensitive to the
effect of offset charge and reduces the effect of charge
fluctuations by using large ratios of $E_{J}/E_{CJ}$. \ A\ large
non-linearity in the Josephson inductance is obtained by biasing
the junction at a current $I$ very close to the critical current.
\ A current bias source can be understood as arising from a loop
inductance with $L\rightarrow \infty $ biased by a flux $\Phi
\rightarrow \infty $ such that $I=\Phi /L$. \ The Hamiltonian is
given by

\begin{equation}
H=E_{CJ}p^{2}-I\varphi _{0}\delta -I_{0}\varphi _{0}\cos \delta
\text{ ,} \label{CBJ-Ham}
\end{equation}
where the gauge invariant phase difference operator $\delta $ is,
apart from the scale factor $\varphi _{0}$, precisely the branch
flux across $C_{J}$. Its conjugate variable is the charge $2ep$ on
that capacitance, a continuous operator. We have thus $\left[
\delta ,p\right] =i$. The variable $\delta $, like the variable
$\phi $ of the RF-SQUID, takes its value on the whole real axis
and its relation with the phase $\theta $ is $\delta \, {\rm{mod}}
2\pi =\theta $ as in our classical analysis of section 4.

\begin{figure}[tbp]
\centering
\includegraphics[width = 0.7\columnwidth]{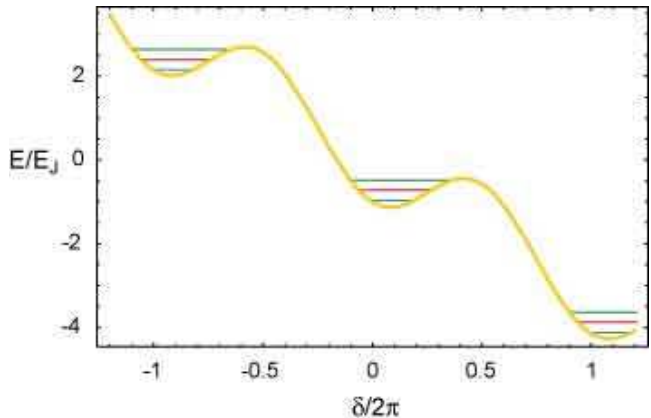}
\caption{Tilted washboard potential of the current-biased
Josephson junction.} \label{fig:fig9-cbjj}
\end{figure}

The potential in the $\delta $ representation is shown in
Fig.~\ref{fig:fig9-cbjj}. It has the shape of a tilted washboard,
with the tilt given by the ratio $I/I_{0}$. When $I$ approaches
$I_{0}$, the phase is $\delta \approx \pi /2$, and in its
vicinity, the potential is very well approximated by the cubic
form
\begin{equation}
U\left( \delta \right) =\varphi _{0}\left( I_{0}-I\right) \left(
\delta -\pi /2\right) -\frac{I_{0}\varphi _{0}}{6}\left( \delta
-\pi /2\right) ^{3} \label{cubic-pot}
\end{equation}
Note that its shape depends critically on the difference
$I_{0}-I$. For $I\lesssim I_{0}$, there is a well with a barrier
height $\Delta U=(2\sqrt{2} /3)I_{0}\varphi _{0}\left(
1-I/I_{0}\right) ^{3/2}$ and the classical oscillation frequency
at the bottom of the well (so-called plasma oscillation) is given
by
\begin{eqnarray*}
\omega _{p} &=&\frac{1}{\sqrt{L_{J}(I)C_{J}}} \\
&=&\frac{1}{\sqrt{L_{J0}C_{J}}}\left[ 1-\left( I/I_{0}\right)
^{2}\right] ^{1/4}
\end{eqnarray*}
Quantum-mechanically, energy levels are found in the well (see
Fig.~\ref{fig:Fig11-Rabi--phasequbit})
\cite{Martinis-Devoret-Clarke} with non-degenerate spacings. \ The
first two levels can be used for qubit states
\cite{Martinis-Urbina}, and have a transition frequency $\omega
_{01}\simeq 0.95 \, \omega _{p}$.

A feature of this qubit circuit is built-in readout, a property
missing from the two previous cases. It is based on the
possibility that states in the cubic potential can tunnel through
the cubic potential barrier into the continuum outside the
barrier. \ Because the tunneling rate increases by a factor of
approximately 500 each time we go from one energy level to the
next, the population of the $\left\vert 1\right\rangle $ qubit
state can be reliably measured by sending a probe signal inducing
a transition from the 1 state to a higher energy state with large
tunneling probability. After tunneling, the particle representing
the phase accelerates down the washboard, a convenient
self-amplification process leading to a voltage $2\Delta /e$
across the junction. Therefore, a finite voltage $V\neq 0$
suddenly appearing across the junction just after the probe signal
implies that the qubit was in state $\left\vert 1\right\rangle $,
whereas $V=0$ implies that the qubit was in state $\left\vert
0\right\rangle $.

In practice, like in the two previous cases, the transition
frequency $\omega _{01}/2\pi $ falls in the 5-20 GHz range. This
frequency is only determined by material properties of the
barrier, since the product $C_{J}$ $L_{J}$ does not depend on
junction area. The number of levels in the well is typically
$\Delta U/\hbar \omega _{p}\approx 4$.

Setting the bias current at a value $I$ and calling $\Delta I$ the
variations of the difference $I-I_{0}$ (originating either in
variations of $I$ or $I_{0})$, the qubit Hamiltonian is given by
\begin{equation}
H_{\rm{qubit}}=\frac{\hbar \omega_{01}}{2}\sigma
_{Z}+\sqrt{\frac{\hbar }{2\omega _{01}C_{J}}}\Delta I(\sigma
_{X}+\chi \sigma _{Z})\text{,} \label{CBJ-qubit-ham}
\end{equation}
where $\chi =\sqrt{\hbar \omega _{01}/3\Delta U}\simeq 1/4$ for
typical operating parameters. In contrast with the flux and phase
qubit circuits, the current-biased Josephson junction does not
have a bias point where the 0$ \rightarrow $1 transition frequency
has a local minimum. The hamiltonian cannot be cast into the
NMR-type form of Eq.~(\ref{box-qubit-ham}). However, a sinusoidal
current signal $\Delta I\left( t\right) \sim \sin \omega _{01}t$
can still produce $\sigma _{X}$\ rotations, whereas a
low-frequency signal produces $\sigma _{Z}$\ operations
\cite{Steffen-Martinis}.

In analogy with the preceding circuits, qubits derived from this
circuit and/or having the same phase potential shape and qubit
properties have been nicknamed \textquotedblleft phase
qubits\textquotedblright\ since the controlled variable is the
phase (the $X$ pseudo-spin direction in
hamiltonian~(\ref{CBJ-qubit-ham})).

\subsection{Tunability versus sensitivity to noise in control parameters}

The reduced two-level hamiltonians
Eqs.~(\ref{box-qubit-ham},\ref{SQUID-qubit-ham}) and
(\ref{CBJ-qubit-ham}) have been tested thoroughly and are now
well-established. They contain the very important parametric
dependence of the coefficient of $\sigma _{X}$, which can be
viewed on one hand as how much the qubit can be tuned by an
external control parameter, and on the other hand as how much it
can be dephased by uncontrolled variations in that parameter. It
is often important to realize that even if the control parameter
has a very stable value at the level of room-temperature
electronics, the noise in the electrical components relaying its
value at the qubit level might be inducing detrimental
fluctuations. An example is the flux through a superconducting
loop, which in principle could be set very precisely by a stable
current in a coil, and which in practice often fluctuates because
of trapped flux motion in the wire of the loop or in nearby
superconducting films. Note that, on the other hand, the two-level
hamiltonian does not contain all the non-linear properties of the
qubit, and how they conflict with its intrinsic noise, a problem
which we discuss in the next subsection.

\subsection{Non-linearity versus sensitivity to intrinsic noise}

The three basic quantum circuit types discussed above illustrate a
general tendency of Josephson qubits. If we try to make the level
structure very non-linear, i.e. $\left\vert \omega _{01}-\omega
_{12}\right\vert \gg \omega _{01}$, we necessarily expose the
system sensitively to at least one type of intrinsic noise. The
flux qubit is contructed to reach a very large non-linearity, but
is also maximally exposed, relatively speaking, to critical
current noise and flux noise. On the other hand, the phase qubit
starts with a relatively small non-linearity and acquires it at
the expense of a precise tuning of the difference between the bias
current and the critical current, and therefore exposes itself
also to the noise in the latter. The Cooper pair box, finally,
acquires non-linearity at the expense of its sensitivity to offset
charge noise. The search for the optimal qubit circuit involves
therefore a detailed knowledge of the relative intensities of the
various sources of noise, and their variations with all the
construction parameters of the qubit, and in particular -- this
point is crucial -- the properties of the materials involved in
the tunnel junction fabrication. Such in-depth knowledge does not
yet exist at the time of this writing and one can only make
educated guesses.

The qubit optimization problem is also further complicated by the
necessity to readout quantum information, which we address just
after reviewing the relationships between the intensity of noise
and the decay rates of quantum information.


\section{Qubit relaxation and decoherence}

A generic quantum state of a qubit can be represented as a unit
vector $\overrightarrow{S}$ pointing on a sphere -- the so-called
Bloch sphere. One distinguishes two broad classes of errors. The
first one corresponds to the tip of the Bloch vector diffusing in
the latitude direction, i.e. along the arc joining the two poles
of the sphere to or away from the North pole. This process is
called energy relaxation or state-mixing. The second class
corresponds to the tip of the Bloch vector diffusing in the
longitude direction, i.e. perpendicularly to the line joining the
two poles. This process is called dephasing or decoherence.

In Appendix~\ref{app:decoherence} we define precisely the
relaxation and decoherence rates and show that they are directly
proportional to the power spectral densities of the noises
entering in the parameters of the hamiltonian of the qubit. More
precisely, we find that the decoherence rate is proportional to
the total spectral density of the quasi-zero-frequency noise in
the qubit Larmor frequency. The relaxation rate, on the other
hand, is proportional to the total spectral density, at the qubit
Larmor frequency, of the noise in the field perpendicular to the
eigenaxis of the qubit.

In principle, the expressions for the relaxation and decoherence
rate could lead to a ranking of the various qubit circuits: from
their reduced spin hamiltonian, one can find with what coefficient
each basic noise source contributes to the various spectral
densities entering in the rates. In the same manner, one could
optimize the various qubit parameters to make them insensitive to
noise, as much as possible. However, before discussing this
question further, we must realize that the readout itself can
provide substantial additional noise sources for the qubit.
Therefore, the design of a qubit circuit that maximizes the number
of coherent gate operations is a subtle optimization problem which
must treat in parallel both the intrinsic noises of the qubit and
the back-action noise of the readout.

\section{Readout of superconducting qubits}

\subsection{Formulation of the readout problem}

We have examined so far the various basic circuits for qubit
implementation and their associated methods to write and
manipulate quantum information. Another important task quantum
circuits must perform is the readout of that information. As we
mentioned earlier, the difficulty of the readout problem is to
open a coupling channel to the qubit for extracting information
without at the same time submitting it to noise.

Ideally, the readout part of the circuit -- referred to in the
following simply as \textquotedblleft readout\textquotedblright\
-- should include both a switch, which defines an
\textquotedblleft OFF\textquotedblright\ and an \textquotedblleft
ON\textquotedblright\ phase, and a state measurement device.
During the OFF phase, where reset and gate operations take place,
the measurement device should be completely decoupled from the
qubit degrees of freedom. During the ON phase, the measurement
device should be maximally coupled to a qubit variable that
distinguishes the 0 and the 1 state. However, this condition is
not sufficient. The back-action of the measurement device during
the ON phase should be weak enough not to relax the qubit
\cite{Devoret-Schoelkopf}.

The readout can be characterized by 4 parameters. The first one
describes the sensitivity of the measuring device while the next
two describe its back-action, factoring in the quality of the
switch:
\begin{itemize}
    \item[i)] the measurement time $\tau _{m}$ defined as the time taken
by the measuring device to reach a signal-to-noise ratio of 1 in
the determination of the state.
    \item[ii)] the energy relaxation rate $\Gamma _{1}^{\rm{ON}}$ of the qubit
in the ON state.
    \item[iii)] the coherence decay rate $\Gamma _{2}^{\rm{OFF}}$ of the qubit
information in the OFF state.
    \item[iv)] the dead time $t_{d}$ needed to reset both the measuring
device and qubit after a measurement. They are usually perturbed
by the energy expenditure associated with producing a signal
strong enough for external detection.
\end{itemize}

Simultaneously minimizing these parameters to improve readout
performance cannot be done without running into conflicts. An
important quantity to optimize is the readout fidelity. By
construction, at the end of the ON phase, the readout should have
reached one of two classical states: 0$_{c}$ and 1$_{c}$, the
outcomes of the measurement process. The latter can be described
by 2 probabilities: the probability $p_{00_{c}}$($p_{11_{c}}$)
that starting from the qubit state $\left\vert 0\right\rangle $
($\left\vert 1\right\rangle $) the measurement yields
0$_{c}$(1$_{c}$ ). The readout fidelity (or discriminating power)
is defined as $F=p_{00c}+p_{11_{c}}-1$. For a measuring device
with a signal-to-noise ratio increasing like the measurement
duration $\tau $, we would have, if back-action could be
neglected, $F=\rm{erf}\left( 2^{-1/2}\sqrt{\tau /\tau _{m}}\right)
$.

\subsection{Requirements and general strategies}

The fidelity and speed of the readout, usually not discussed in
the context of quantum algorithms because they enter marginally in
the evaluation of their complexity, are actually key to
experiments studying the coherence properties of qubits and gates.
A very fast and sensitive readout will gather at a rapid pace
information on the imperfections and drifts of qubit parameters,
thereby allowing the experimenter to design fabrication strategies
to fight them during the construction or even correct them in real
time.

We are thus mostly interested in \textquotedblleft
single-shot\textquotedblright\ readouts \cite{Devoret-Schoelkopf},
for which $F$ is of order unity, as opposed to schemes in which a
weak measurement is performed continuously \cite{Averin-Korotkov}.
If $F\ll 1$, of order $F^{-2}$ identical preparation and readout
cycles need to be performed to access the state of the qubit. The
condition for \textquotedblleft single-shot\textquotedblright\
operation is

\begin{equation*}
\Gamma _{1}^{\rm{ON}}\tau _{m}<1
\end{equation*}

The speed of the readout, determined both by $\tau _{m}$ and
$t_{d}$, should be sufficiently fast to allow a complete
characterization of all the properties of the qubit before any
drift in parameters occurs. With sufficient speed, the automatic
correction of these drits in real time using feedback will be
possible.

Rapidly pulsing the readout on and off with a large decoupling
amplitude such that

\begin{equation*}
\Gamma _{2}^{\rm{OFF}}T_{2}\ll 1
\end{equation*}
requires a fast, strongly non-linear element, which is provided by
one or more auxiliary Josephson junctions. Decoupling the qubit
from the readout in the OFF phase requires balancing the circuit
in the manner of a Wheatstone bridge, with the readout input
variable and the qubit variable corresponding to 2 orthogonal
electrical degrees of freedom. Finally, to be as complete as
possible even in presence of small asymmetries, the decoupling
also requires an impedance mismatch between the qubit and the
dissipative degrees of freedom of the readout. In the next
subsection, we discuss how these general ideas have been
implemented in 2nd generation quantum circuits. The first three
examples we have chosen involve a readout circuit which is
built-in the qubit itself to provide maximal coupling during the
ON phase, as well as a decoupling scheme which has proven
effective for obtaining long decoherence times. The last example
incorporates a novel dispersive readout.

\subsection{Phase qubit: tunneling readout with a DC-SQUID on-chip amplifier.}

\begin{figure}[tbp]
\centering
\includegraphics[width = 0.4\columnwidth]{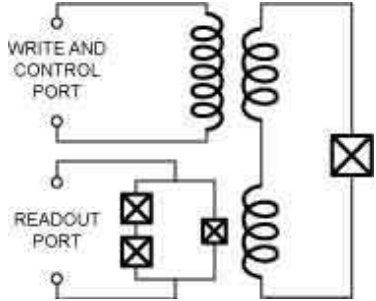}
\caption{Phase qubit implemented with a Josephson junction in a
high-inductance superconducting loop biased with a flux
sufficiently large that the phase across the junction sees a
potential analogous to that found for the current-biased junction.
The readout part of the circuit is an asymmetric hysteretic SQUID
which is completely decoupled from the qubit in the OFF phase.
Isolation of the qubit both from the readout and control port is
obtained through impedance mismatch of transformers.}
\label{fig:fig10-nist-new-phasequbit}
\end{figure}

\begin{figure}[tbp]
\centering
\includegraphics[width = 0.70\columnwidth]{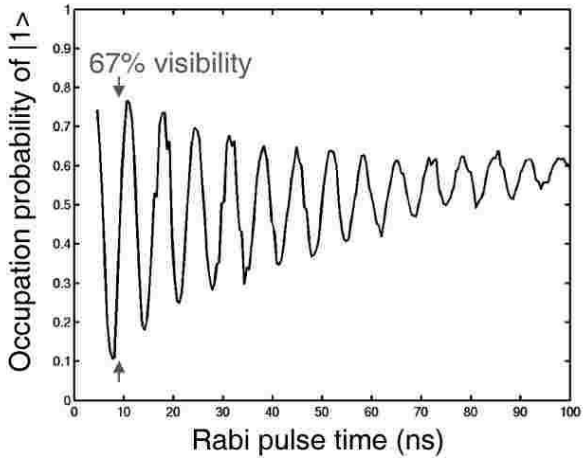}
\caption{Rabi oscillations observed for the qubit of Fig. 10.}
\label{fig:Fig11-Rabi--phasequbit}
\end{figure}

The simplest example of a readout is provided by a system derived
from the phase qubit (See
Fig.~\ref{fig:fig10-nist-new-phasequbit}).
In the phase qubit, the levels in the cubic potential are
metastable and decay in the continuum, with level $n+1$ having
roughly a decay rate $\Gamma _{n+1}$ 500 times faster than the
decay $\Gamma _{n}$ of level $n$. This strong level number
dependence of the decay rate leads naturally to the following
readout scheme: when readout needs to be performed, a microwave
pulse at the transition frequency $\omega _{12}$ (or better at
$\omega _{13})$ transfers the eventual population of level $1$
into level $2$, the latter decaying rapidly into the continuum,
where it subsequently loses energy by friction and falls into the
bottom state of the next corrugation of the potential (because the
qubit junction is actually in a superconducting loop of large but
finite inductance, the bottom of this next corrugation is in fact
the absolute minimum of the potential and the particle
representing the system can stay an infinitely long time there).
Thus, at the end of the readout pulse, the sytem has either
decayed out of the cubic well (readout state 1$_{c}$) if the qubit
was in the $\left\vert 1\right\rangle $ state or remained in the
cubic well (readout state 0$_{c}$) if the qubit was in the
$\left\vert 0\right\rangle $ state. The DC-SQUID amplifier is
sensitive enough to detect the change in flux accompanying the
exit of the cubic well, but the problem is to avoid sending the
back-action noise of its stabilizing resistor into the qubit
circuit. The solution to this problem involves balancing the SQUID
loop in such a way, that for readout state 0$_{c}$, the small
signal gain of the SQUID is zero, whereas for readout state
1$_{c}$, the small signal gain is non-zero \cite{Martinis-TLS}.
This signal dependent gain is obtained by having 2 junctions in
one arm of the SQUID whose total Josephson inductance equals that
of the unique junction in the other arm. Finally, a large
impedance mismatch between the SQUID\ and the qubit is obtained by
a transformer. The fidelity of such readout is remarkable: 95\%
has been demonstrated. In Fig.~\ref{fig:Fig11-Rabi--phasequbit},
we show the result of a measurement of Rabi oscillations with such
qubit plus readout.

\subsection{Cooper-pair box with non-linear inductive readout: the
\textquotedblleft Quantronium\textquotedblright\ circuit}

\begin{figure}[tbp]
\centering
\includegraphics[width = 0.7\columnwidth]{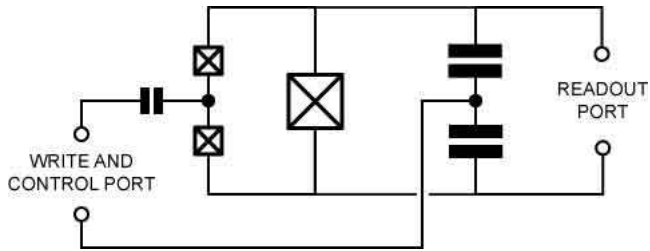}
\caption{\textquotedblleft Quantronium\textquotedblright\ circuit
consisting of a Cooper pair box with a non-linear inductive
readout. A Wheatstone bridge configuration decouples qubit and
readout variables when readout is OFF. Impedance mismatch
isolation is also provided by additional capacitance in parallel
with readout junction.} \label{fig:fig12-quantronium}
\end{figure}

The Cooper-pair box needs to be operated at its \textquotedblleft
sweet spot\textquotedblright\ (degeneracy point) where the
transition frequency is to first order insensitive to offset
charge fluctuations. The \textquotedblleft
Quantronium\textquotedblright\ circuit presented in
Fig.~\ref{fig:fig12-quantronium} is a 3-junction bridge
configuration with two small junctions defining a Cooper box
island, and thus a charge-like qubit which is coupled capacitively
to the write and control port (high-impedance port). There is also
a large third junction, which provides a non-linear inductive
coupling to the read port. When the read port current $I$ is zero,
and the flux through the qubit loop is zero, noise coming from the
read port is decoupled from the qubit, provided that the two small
junctions are identical both in critical current and capacitance.
When $I$ is non-zero, the junction bridge is out of balance and
the state of the qubit influences the effective non-linear
inductance seen from the read port. A further protection of the
impedance mismatch type is obtained by a shunt capacitor across
the large junction: at the resonance frequency of the non-linear
resonator formed by the large junction and the external
capacitance $C$, the differential mode of the circuit involved in
the readout presents an impedance of the order of an ohm, a
substantial decoupling from the 50 $\Omega$ transmission line
carrying information to the amplifier stage. The readout protocol
involves a DC pulse \cite{Cottet-Physica, Vion-Science} or an RF
pulse \cite{Siddiqi} stimulation of the readout mode. The response
is bimodal, each mode corresponding to a state of the qubit.
Although the theoretical fidelity of the DC readout can attain
95\%, only a maximum of 40\% has been obtained so far. The cause
of this discrepancy is still under investigation.

\begin{figure}[!h] \centering
\includegraphics[width = 0.6\columnwidth]{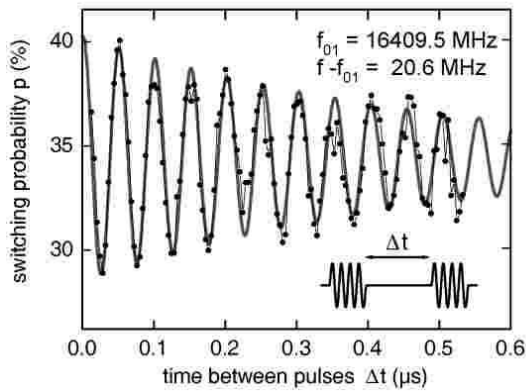}
\caption{Measurement of Ramsey fringes for the Quantronium. Two
$\protect\pi /2$ pulses separated by a variable delay are applied
to the qubit before measurement. The frequency of the pulse is
slightly detuned from the transition frequency to provide a
stroboscopic measurement of the Larmor precession of the qubit.}
\label{fig:fig13-ramsey_quantronium}
\end{figure}

In Fig.~\ref{fig:fig13-ramsey_quantronium} we show the result of a
Ramsey fringe experiment demonstrating that the coherence quality
factor of the quantronium can reach 25 000 at the sweet spot \cite
{Vion-Science}.
By studying the degradation of the qubit
absorption line and of the Ramsey fringes as one moves away from
the sweet spot, it has been possible to show that the residual
decoherence is limited by offset charge noise and by flux noise
\cite{Vion-Fort}. In principle, the influence of these noises
could be further reduced by a better optimization of the qubit
design and parameters. In particular, the operation of the box can
tolerate ratios of $E_{J}/E_{C}$ around 4 where the sensitivity to
offset charge is exponentially reduced and where the non-linearity
is still of order 15\%. The quantronium circuit has so far the
best coherence quality factor. We believe this is due to the fact
that critical current noise, one dominant intrinsic source of
noise, affects this qubit far less than the others, relatively
speaking, as can be deduced from the qubit hamiltonians of section
6.

\subsection{3-junction flux qubit with built-in readout}

\begin{figure}[tbp]
\centering
\includegraphics[width = 0.6\columnwidth]{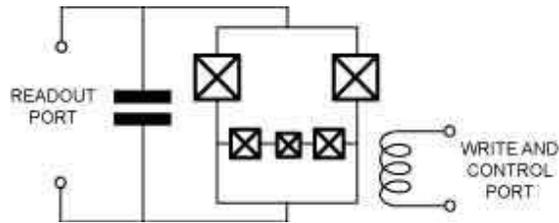}
\caption{Three-junction flux qubit with a non-linear inductive
readout. The medium-size junctions play the role of an inductor.
Bridge configuration for nulling out back-action of readout is
also employed here, as well as impedance mismatch provided by
additional capacitance.} \label{fig:fig14-delft-fluxqubit}
\end{figure}

\begin{figure}[tbp]
\centering
\includegraphics[width = 0.7\columnwidth]{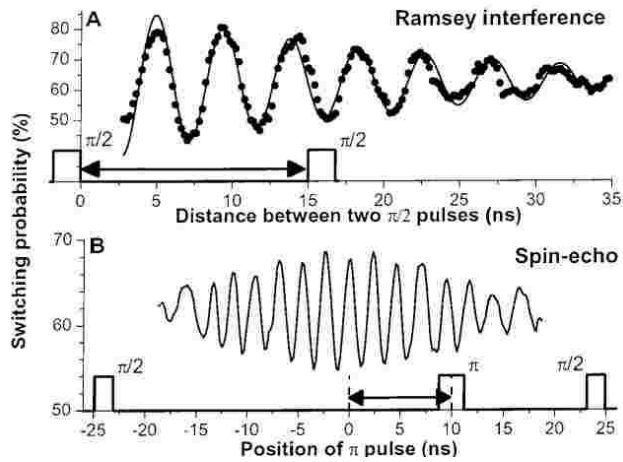}
\caption{Ramsey fringes obtained for qubit of Fig. 14.}
\label{fig:fig15-ramsey-flux-qubit}
\end{figure}

Fig.~\ref{fig:fig14-delft-fluxqubit} shows a third example of
built-in readout, this time for a flux-like qubit. The qubit by
itself involves 3 junctions in a loop, the larger two of the
junctions playing the role of the loop inductance in the basic
RF-SQUID \cite{Mooij-Nak}. The advantage of this configuration is
to reduce the sensitivity of the qubit to external flux
variations. The readout part of the circuit involves 2 other
junctions forming a hysteretic DC-SQUID whose offset flux depends
on the qubit flux state. The critical current of this DC-SQUID has
been probed by a DC pulse, but an RF pulse could be applied as in
another flux readout. Similarly to the two previous cases, the
readout states 1$_{c}$ and 0$_{c}$, which here correspond to the
DC-SQUID having switched or not, map very well the qubit states
$\left\vert 1\right\rangle $ and $\left\vert 0\right\rangle $,
with a fidelity better than 60\%. Here also, a bridge technique
orthogonalizes the readout mode, which is the common mode of the
DC-SQUID, and the qubit mode, which is coupled to the loop of the
DC-SQUID. External capacitors provide additional protection
through impedance mismatch. Fig.~\ref{fig:fig15-ramsey-flux-qubit}
shows Ramsey oscillations obtained with this system.

\subsection{Avoiding on-chip dissipation with dispersive readout schemes}
All the circuits above include an on-chip amplification scheme
producing high-level signals which can be read directly by
high-temperature low-noise electronics. In the second and third
examples, these signals lead to non-equilibrium quasiparticle
excitations being produced in the near vicinity of the qubit
junctions.
More generally, one can legitimately worry that large energy
dissipation on the chip itself will lead to an increase of the
noises discussed in section 5.2. A broad class of new readout
schemes addresses this question \cite{Circuit-QED, Siddiqi,
Lupascu-Mooij}. They are based on a purely dispersive measurement
of a qubit susceptibility (capacitive or inductive). A probe
signal is sent to the qubit. The signal is coupled to a qubit
variable whose average value is identical in the 2 qubit states
(for instance, in the capacitive susceptibility, the variable is
the island charge in the charge qubit at the degeneracy point).
However, the susceptibility, which is the derivative of the qubit
variable with respect to the probe, differs from one qubit state
to the other. The resulting state-dependent phase shift of the
reflected signal is thus amplified by a linear low-temperature
amplifier and finally discriminated at high temperature against an
adequately chosen threshold. In addition to being very thrifty in
terms of energy being dissipated on chip, these new schemes also
provide a further natural decoupling action: when the probe signal
is off, the back-action of the amplifier is also completely shut
off. Finally, the interrogation of the qubit in a frequency band
excluding zero facilitates the design of very efficient filters.

\subsection{Cooper-pair box with circuit quantum electrodynamics readout}
\begin{figure}[tbp]
\centering
\includegraphics[width = 0.6\columnwidth]{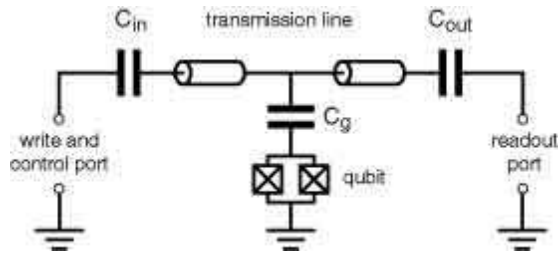}
\caption{Circuit quantum electrodynamics architecture for readout
of a Cooper pair box. The Cooper pair box is coupled capacitively
($C_{\rm{g}}$) to a transmission line resonator with capacitively
coupled input ($C_{\rm{in}}$) and output ($C_{\rm{out}}$) ports.
The input port is used to control the qubit state using microwave
pulses and to apply the readout microwave which, after
transmission through the circuit, is detected at the readout
port.} \label{fig:Fig16-CircuitQEDarchitecture}
\end{figure}

\begin{figure}[tbp]
\centering
\includegraphics[width = 0.5\columnwidth]{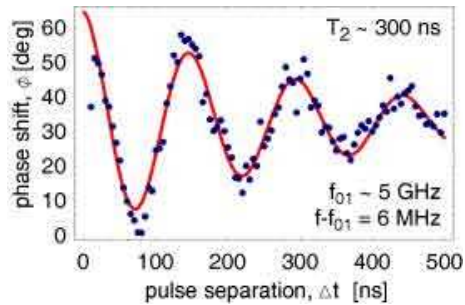}
\caption{Ramsey oscillations observed in the phase of the
transmitted probe beam in the circuit QED readout architecture.
The drive frequency is detuned by $6 \, \rm{MHz}$ from the qubit
transition frequency. A dephasing time of $T_2 \sim 300 \,
\rm{ns}$ is observed.} \label{fig:Fig16-CircuitQEDRamsey}
\end{figure}

As a representative example of the novel class of dispersive qubit
readout schemes we discuss  the circuit quantum electrodynamics
architecture \cite{Circuit-QED} applied to the readout of charge
qubits in some more detail. In this approach a split Cooper pair
box is capacitively coupled to a single mode of the
electromagnetic field contained in a high quality on-chip
transmission line resonator, see
Fig.~\ref{fig:Fig16-CircuitQEDarchitecture}.
Exciting the resonator at its full wave resonance frequency, the
electric field has an antinode at the center of the resonator to
which the qubit is strongly coupled by a large capacitance
$C_{\rm{g}}$. It has been demonstrated that the coupling can be
made so large that a single photon in the transmission line can
resonantly drive Rabi oscillations in the Cooper pair box at
frequencies in excess of $10 \, \rm{MHz}$ \cite{wallraff}. This
rate of coherent exchange of a single excitation between the
Cooper pair box and the resonator is much larger than the rates at
which the Cooper pair box decoheres or the photon gets lost from
the resonator. When the qubit transition frequency is detuned form
the resonator, the mutual strong coupling gives rise to a qubit
state-dependent frequency shift in the resonator transition
frequency. This frequency shift can be used to perform a quantum
non-demolition (QND) measurement of the qubit state by measuring
the amplitude and phase of a probe microwave transmitted through
the resonator at its bare resonance frequency
\cite{wallraff,schuster}. In this dispersive scheme, the
susceptibility of the Cooper pair box which is maximal at charge
degeneracy (the optimal bias point) is measured. The measurement
is performed without dissipating any power in the circuit and thus
minimizing the back-action. The only source of back-action in this
QND measurement is the dephasing induced by the fluctuating
ac-Stark shift in the qubit level separation due to the photon
shot noise in the microwave readout beam \cite{schuster}. In the
detuned case, the resonator also enhances the qubit radiative
lifetime by providing an impedance transformation which
effectively filters out the noise of the electromagnetic
environment at the qubit transition frequency. Using this
architecture, it has been demonstrated that the quantum state of
the Cooper pair box can be manipulated by applying microwave
pulses at the qubit transition frequency to the write and control
port of the circuit QED system. A measurement of Ramsey
oscillations, showing a coherence time of $T_2 \sim 300 \,
\rm{ns}$ at the optimal bias point, under a weak continuous probe
beam is shown in Fig.~\ref{fig:Fig16-CircuitQEDRamsey}.

\section{Coupling superconducting qubits}

A priori, 3 types of coupling schemes can be envisioned:

\begin{itemize}
    \item[a)]In the first type, the transition frequency of the qubits are
all equal and the coupling between any pair is switched on using
one or several junctions as non-linear elements \cite{Averin,
Blais}.
    \item[b)] In the second type, the couplings are fixed, but the transition
frequencies of a pair of qubits, originally detuned, are brought
on resonance when the coupling between them needs to be turned on
\cite{Nak-Coupled-qubits, Majer-Mooij, Wellstood}.
    \item[c)] In the third type, which bears close resemblance to the methods used in
NMR \cite{Nielsen_Chuang}, the couplings and the resonance
frequencies of the qubits remain fixed, the qubits being always
detuned. Being off-diagonal, the coupling elements have negligible
action on the qubits. However, when a strong microwave field is
applied to the target and control qubits at their resonant
frequency and for an appropriate amplitude, they become in
\textquotedblleft speaking terms\textquotedblright\ for the
exchange of energy quanta and gate action can take place
\cite{Rigetti}.
\end{itemize}

So far only scheme b) has been tested experimentally.

The advantage of schemes b) and c) is that they work with purely
passive reactive elements like capacitors and inductors which
should remain very stable as a function of time and which also
should present very little high-frequency noise. In a way, we must
design quantum integrated circuits in the manner that vacuum tube
radios were designed in the 50's: only 6 tubes were used for a
complete heterodyne radio set, including the power supply.
Nowadays several hundreds of transistors are used in a radio or
any hi-fi system. In that ancient era of classical electronics,
linear elements like capacitors, inductors or resistors were
\textquotedblleft free\textquotedblright\ because they were
relatively reliable whereas tubes could break down easily. We have
to follow a similar path in quantum integrated circuit, the
reliability issues having become noise minimization issues.

\section{Can coherence be improved with better materials?}

Up to now, we have discussed how, given the power spectral
densities of the noises $\Delta Q_{r}$, $\Delta E_{C}$ and $\Delta
E_{J}$, we could design a qubit equipped with control, readout and
coupling circuits. It is worthwhile to ask at this point if we
could improve the material properties to improve the coherence of
the qubit, assuming all other problems like noise in the control
channels and the back-action of the readout have been solved. A
model put forward by one of us (JMM) and collaborators shed some
light on the direction one would follow to answer this question.
The $1/f$ spectrum of the materials noises suggests that they all
originate from 2-level fluctuators in the amorphous alumina tunnel
layer of the junction itself, or its close vicinity. The substrate
or the surface of the superconducting films are also suspect in
the case of $\Delta Q_{r}$ and $\Delta E_{C}$ but their influence
would be relatively weaker and we ignore them for simplicity.
These two-level systems are supposed to be randomly distributed
positional degrees of freedom $\xi _{i}$ with effective spin-1/2
properties, for instance an impurity atom tunneling between two
adjacent potential wells. Each two-level system is in principle
characterized by 3 parameters: the energy splitting $\hbar \omega
_{i}$, and the two coefficients $\alpha _{i}$ and $\beta _{i}$ of
the Pauli matrix representation of $\xi _{i}=\alpha _{i}\sigma
_{iz}+\beta _{i}\sigma _{ix}$ ($z$ here is by definition the
energy eigenbasis). The random nature of the problem leads us to
suppose that $\alpha _{i}$ and $\beta _{i}$ are both Gaussian
random variables with the same standard deviation $\rho _{i}$. By
carrying a charge, the thermal and quantum motion of $\xi _{i}$
can contribute to $\Delta Q_{r}=\sum_{i}q_{i}\xi _{i}$ and $\Delta
E_{C}=\sum_{i}c_{i}{\beta _{i}^{2}}/{\omega _{i}} \, \sigma
_{iz}$. Likewise, by modifying the transmission of a tunneling
channel in its vicinity, the motion of $\xi _{i}$ can contribute
to $\Delta E_{J}=\sum_{i}g_{i}\xi _{i}$. We can further suppose
that the quality of the material of the junction is simply
characterized by a few numbers. The essential one is the density
$\nu $ of the transition frequencies $\omega _{i}$ in frequency
space and in real space, assuming a $\omega ^{-1}$ distribution
(this is necessary to explain the $1/f$ behavior) and a uniform
spatial distribution on the surface of the junction. Recent
experiments indicate that the parameter $\nu $ is of order $10^{5}
\, \mathrm{\mu m}^{-2}(\mathrm{decade})^{-1}$. Then, assuming a
universal value for $\rho = \langle \rho_i \rangle$ which is
independent of frequency, only one coefficient is needed per
noise, namely, the average modulation efficiency of each
fluctuator. Such analysis provides a common language for
describing various experiments probing the dependence of
decoherence on the material of the junction. Once the influence of
the junction fabrication parameters (oxidation pressure and
temperature, impurity contents, and so on) on these noise
intensities will be known, it will be possible to devise optimized
fabrication procedures, in the same way perhaps as the $1/f$ noise
in C-MOS transistors has been reduced by careful material studies.

\section{Concluding remarks and perspectives}

The logical thread through this review of superconducting qubits
has been the question \textquotedblleft What is the best qubit
design?\textquotedblright . Because some crucial experimental data
is still missing, we unfortunately, at present, cannot conclude by
giving a definitive answer to this complex optimization problem.

Yet, a lot has already been achieved, and superconducting qubits
are becoming serious competitors of trapped ions and atoms. The
following properties of quantum circuits have been demonstrated:

\begin{itemize}
    \item[a)] Coherence quality factors $Q_{\varphi }=T_{\varphi }\omega _{01}$ can
attain at least 2$\times 10^{4}.$
    \item[b)] Readout and reset fidelity can be greater than 95\%.
    \item[c)] All states on the Bloch sphere can be addressed.
    \item[d)] Spin echo techniques can null out low frequency drift of offset charges.
    \item[e)] Two qubits can be coupled and RF pulses can implement gate operation.
    \item[f)] A qubit can be fabricated using only optical lithography techniques.
\end{itemize}

The major problem we are facing is that these various results have
not been obtained at the same time in the same circuit, although
succesful design elements in one have often been incorporated into
the next generation of others. The complete optimization of the
single qubit plus readout has not been achieved yet. However, we
have presented in this review the elements of a systematic
methodology resolving the various conflicts that are generated by
all the different requirements. Our opinion is that, once noise
sources are better characterized, an appropriate combination of
all the known circuit design strategies for improving coherence,
as well as the understanding of optimal tunnel layer growth
conditions for lowering the intrinsic noise of Josephson
junctions, should lead us to reach the 1-qubit and 2-qubit
coherence levels needed for error correction \cite{Preskill}.
Along the way, good medium term targets to test overall progress
on the simultaneous fronts of qubit coherence, readout and gate
coupling are the measurement of Bell 's inequality violation or
the implementation of the Deutsch-Josza algorithm, both of which
requiring the simultaneous satisfaction of properties a)-e).

\subsection*{Acknowlegements}
The authors have greatly benefited from discussions with I.
Chuang, D. Esteve, S. Girvin, S. Lloyd, H. Mooij, R. Schoelkopf,
I. Siddiqi, C. Urbina and D. Vion. They would like also to thank
the participants of the Les Houches Summer School on Quantum
Information Processing and Entanglement held in 2003 for useful
exchanges. Finally, funding from ARDA/ARO and the Keck Fundation
is gratefully acknowledged.

\textit{\bigskip }


\bigskip \bigskip

\appendix

{\LARGE \noindent \textbf{Appendix}}

\bigskip

\section{Quantum circuit theory}
\label{app:quantumcircuittheory}

The problem we are addressing in this section is, given a
superconducting circuit made up of capacitors, inductors and
Josephson junctions, how to systematically write its quantum
hamiltonian, the generating function from which the quantum
dynamics of the circuit can be obtained. This problem has been
considered first by Yurke and Denker\cite{Yurke-Denker} in a
seminal paper and analyzed in further details by
Devoret\cite{Devoret-in-Reynaud}. We will only summarize here the
results needed for this review.

The circuit is given as a set of branches, which can be
capacitors, inductors or Josephson tunnel elements, connected at
nodes. Several independent paths formed by a succession of
branches can be found between nodes. The circuit can therefore
contain one or several loops. It is important to note that a
circuit has not one hamiltonian but many, each one depending on a
particular representation. We are describing here one particular
type of representation, which is usually well adapted to circuits
containing Josephson junctions. Like in classical circuit theory,
a set of independent current and voltages has to be found for a
particular representation. We start by associating to each branch
$b$ of the circuit, the current $i_{b}$ flowing through it and the
voltage $v_{b}$ across it (a convention has to be made first on
the direction of the branches). Kirchhoff's laws impose relations
among branch variables and some of them are redundant. The
following procedure is used to eliminate redundant branches: one
node of the circuit is first chosen as ground. Then from the
ground, a loop-free set of branches called spanning tree is
selected. The rule behind the selection of the spanning tree is
the following: each node of the circuit must be linked to the
ground by one and only one path belonging to the tree. In general,
inductors (linear or non-linear) are preferred as branches of the
tree but this is not necessary. Once the spanning tree is chosen
(note that we still have many possibilities for this tree), we can
associate to each node a \textquotedblleft node
voltage\textquotedblright\ $v_{n}$ which is the algebraic sum of
the voltages along the branches between ground and the node. The
conjugate \textquotedblleft node current\textquotedblright\
$i_{n}$ is the algebraic sum of all currents flowing to the node
through capacitors only. The dynamical variables appearing in the
hamiltonian of the circuit are the node fluxes and node charges
defined as

\begin{eqnarray*}
\phi _{n} &=&\int_{-\infty }^{t}v\left( t_{1}\right) dt_{1} \\
q_{n} &=&\int_{-\infty }^{t}i\left( t_{1}\right) dt_{1}
\end{eqnarray*}

Using Kirchhoff's laws, it is possible to express the flux and the
charge of each branch as a linear combination of all the node
fluxes and charges, respectively. In this inversion procedure, the
total flux through loops imposed by external flux bias sources and
polarisation charges of nodes imposed by charge bias sources,
appear.

If we now sum the energies of all branches of the circuit
expressed in terms of node flux and charges, we will obtain the
hamiltonian of the circuit corresponding to the representation
associated with the particular spanning tree. In this hamiltonian,
capacitor energies behave like kinetic terms while the inductor
energies behave as potential terms. The hamiltonian of the $LC$
circuit written in section 2 is an elementary example of this
procedure.

Once the hamiltonian is obtained it is easy get its quantum
version by replacing all the node fluxes and charges by their
quantum operator equivalent. The flux and charge of a node have a
commutator given by $i\hbar $, like the position and momentum of a
particle:

\begin{eqnarray*}
\phi &\rightarrow &\hat{\phi} \\
q &\rightarrow &\hat{q} \\
\left[ \hat{\phi},\hat{q}\right] &=&i\hbar
\end{eqnarray*}

One can also show that the flux and charge operators corresponding
to a branch share the same commutation relation. Note that for the
special case of the Josephson element, the phase $\hat{\theta}$
and Cooper pair number $\hat{N}$ , which are its dimensionless
electric variables, have the property:

\begin{equation*}
\left[ \hat{\theta},\hat{N}\right] =i
\end{equation*}

In the so-called charge basis, we have

\begin{eqnarray*}
\hat{N} &=&\sum_{N}N\left| N\right\rangle \left\langle N\right| \\
\cos \hat{\theta} &=&\frac{1}{2}\sum_{N}\left( \left|
N\right\rangle \left\langle N+1\right| +\left| N+\right\rangle
\left\langle N\right| \right)
\end{eqnarray*}

while in the so-called phase basis, we have

\begin{equation*}
\hat{N}=\left| \theta \right\rangle \frac{\partial }{i\partial
}\left\langle \theta \right|
\end{equation*}

Note that since the Cooper pair number $\hat{N}$ is an operator
with integer eigenvalues, its conjugate variable $\hat{\theta}$,
has eigenvalues behaving like angles, i.e. they are defined only
modulo $2\pi $.

In this review, outside this appendix, we have dropped the hat on
operators for simplicity.

\bigskip

\section{Eigenenergies and eigenfunctions of the Cooper pair box\medskip }
\label{app:CPBeigenenergies}

From Appendix~\ref{app:quantumcircuittheory}, it easy to see that
the hamiltonian of the Cooper pair box leads to the Schrodinger
equation

\begin{equation*}
\left[ E_{C}\left( \frac{\partial }{i\partial \theta
}-N_{g}\right) ^{2}-E_{J}\cos \theta \right] \Psi _{k}\left(
\theta \right) =E_{k}\Psi _{k}\left( \theta \right)
\end{equation*}

The functions $\Psi _{k}\left( \theta \right)
\mathrm{e}^{-iN_{g}}$ and energies $E_{k}$ are solutions of the
Mathieu equation and can be found with arbitrary precision for all
values of the parameters $N_{g}$ and $E_{J}/E_{C}
$\cite{Cottet-Thesis}. For instance, using the program
Mathematica, we find

\begin{eqnarray*}
E_{k} &=&E_{C}\mathcal{M}_{A}\left[ k+1-(k+1)\,{\rm{mod}}2+2N_{g}(-1)^{k},-2E_{J}/E_{C}\right] \\
\Psi _{k}\left( \theta \right) &=&\frac{e^{iN_{g}\theta
}}{\sqrt{2\pi }} \left\{ \mathcal{M}_{C}\left[
\frac{4E_{k}}{E_{C}},\frac{-2E_{J}}{E_{C}}, \frac{\theta
}{2}\right] +i(-1)^{k+1}\mathcal{M}_{S}\left[ \frac{4E_{k}}{
E_{C}},\frac{-2E_{J}}{E_{C}},\frac{\theta }{2}\right] \right\}
\end{eqnarray*}

where
$\mathcal{M}_{A}(r,q)=\mathtt{MathieuCharacteristicA}$\texttt{[r,q],}

$\mathcal{M}_{C}\left( a,q,z\right)
=\mathtt{MathieuC}$\texttt{[a,q,z],}

$\mathcal{M}_{S}\left( a,q,z\right)
=\mathtt{MathieuS}$\texttt{[a,q,z].}

\bigskip

\section{Relaxation and decoherence rates for a qubit}
\label{app:decoherence}

\subsection{Definition of the rates}

We start by introducing the spin eigenreference frame $\hat{z}$,
$\hat{x}$ and $\hat{y}$ consisting of the unit vector along the
eigenaxis and the associated orthogonal unit vectors ($\hat{x}$ is
in the $XZ$ plane). For instance, for the Cooper pair box, we find
that $\hat{z}=\cos \alpha \hat{Z} +\sin \alpha \hat{X}$, with
$\tan \alpha =2E_{C}\left( N_{g}-1/2\right) /E_{J}$, while
$\hat{x}=-\sin \alpha \hat{Z}+\cos \alpha \hat{X}$.

Starting with $\overrightarrow{S}$ pointing along $\hat{x}$ at
time $t=0$, the dynamics of the Bloch vector in absence of
relaxation or decoherence is

\begin{equation*}
\overrightarrow{S}_{0}\left( t\right) =\cos \left( \omega
_{01}\right) \hat{x }+\sin \left( \omega _{01}\right) \hat{y}
\end{equation*}

In presence of relaxation and decoherence, the Bloch vector will
deviate from $\overrightarrow{S}_{0}\left( t\right) $ and will
reach eventually the equilibrium value $S_{z}^{eq}\hat{z}$, where
$S_{z}^{eq}=\tanh \left({\hbar \omega _{01}}/{2k_{B}T}\right)$.

We define the relaxation and decoherence rates as

\begin{eqnarray*}
\Gamma _{1} &=&\underset{t\rightarrow \infty }{\lim }\frac{\ln
\left\langle
S_{z}\left( t\right) -S_{z}^{eq}\right\rangle }{t} \\
\Gamma _{\phi } &=&\underset{t\rightarrow \infty }{\lim }\frac{\ln
\left[ \frac{\left\langle \overrightarrow{S}\left( t\right)
.\overrightarrow{S}_{0}\left( t\right) \right\rangle }{\left|
\overrightarrow{S}\left( t\right) -S_{z}^{eq}\hat{z}\right|
}\right] }{t}
\end{eqnarray*}

Note that these rates have both a useful and rigorous meaning only
if the evolution of the components of the average Bloch vector
follows, after a negligibly short settling time, an exponential
decay. The $\Gamma _{1}$ and $ \Gamma _{\phi }$ rates are related
to the NMR spin relaxation times $T_{1}$ and $T_{2}$
\cite{Abragam} by

\begin{eqnarray*}
T_{1} &=&\Gamma _{1}^{-1} \\
T_{2} &=&\left( \Gamma _{\phi }+\Gamma _{1}/2\right) ^{-1}
\end{eqnarray*}

The $T_{2}$ time can be seen as the net decay time of quantum
information, including the influence of both relaxation and
dephasing processes. In our discussion of superconducting qubits,
we must separate the contribution of the two types of processes
since their physical origin is in general very different and
cannot rely on the $T_{2}$ time alone.

\subsection{Expressions for the rates}

The relaxation process can be seen as resulting from unwanted
transitions between the two eigenstate of the qubit induced by
fluctuations in the effective fields along the $x$ and $y$ axes.
Introducing the power spectral density of this field, one can
demonstrate from Fermi's Golden Rule that, for perturbative
fluctuations,

\begin{equation*}
\Gamma _{1}=\frac{S_{x}\left( \omega _{01}\right) +S_{y}\left(
\omega _{01}\right) }{\hbar ^{2}}
\end{equation*}

Taking the case of the Cooper pair box as an example, we find that
$S_{y}\left( \omega _{01}\right) =0$ and that

\begin{equation*}
S_{x}\left( \omega \right) =\int_{-\infty }^{+\infty
}dt\mathrm{e}^{i\omega t}\left\langle A\left( t\right) A\left(
0\right) \right\rangle +\left\langle B\left( t\right) B\left(
0\right) \right\rangle
\end{equation*}
where

\begin{eqnarray*}
A\left( t\right) &=&\frac{\Delta E_{J}\left( t\right)
E_{el}}{2\sqrt{
E_{J}^{2}+E_{el}^{2}}} \\
B\left( t\right) &=&\frac{E_{J}\Delta E_{el}\left( t\right)
}{2\sqrt{
E_{J}^{2}+E_{el}^{2}}} \\
E_{el} &=&2E_{C}\left( N_{g}-1/2\right)
\end{eqnarray*}
Since the fluctuations $\Delta E_{el}\left( t\right) $ can be
related to the impedance of the environment of the
box\cite{Bouchiat-etal, Schoen, Schoelkopf}, an order of magnitude
estimate of the relaxation rate can be performed, and is in rough
agreement with observations \cite{Schoelkopf-Delsing,
Vion-Science}.

The decoherence process, on the other hand, is induced by
fluctuations in the effective field along the eigenaxis $z$. If
these fluctuations are Gaussian, with a white noise spectral
density up to frequencies of order several $\Gamma _{\phi }$
(which is often not the case because of the presence of 1/f noise)
we have

\begin{equation*}
\Gamma _{\phi }=\frac{S_{z}\left( \omega \simeq 0\right) }{\hbar
^{2}}
\end{equation*}
In presence of a low frequency noise with an 1/f behavior, the
formula is more complicated\cite{Martinis-noise}. If the
environment producing the low frequency noise consists of many
degrees of freedom, each of which is very weakly coupled to the
qubit, then one is in presence of classical dephasing which, if
slow enough, can in principle be fought using echo techniques. If,
on the other hand, only a few degrees of freedom like magnetic
spins or glassy two-level systems are dominating the low frequency
dynamics, dephasing is quantum and not correctable, unless the
transition frequencies of these few perturbing degrees of freedom
is itself very stable.

\end{document}